\documentclass[journal,10pt]{IEEEtran}
\usepackage[utf8]{inputenc}
\usepackage{amsfonts}
\usepackage{amssymb}
\usepackage{amsmath}
\usepackage{microtype}
\usepackage{graphicx}
\usepackage{booktabs}
\usepackage{hyperref}
\usepackage{multirow}
\usepackage{footnote}
\usepackage{tablefootnote}
\usepackage{textcomp, gensymb} 
\usepackage{commath} 
\usepackage{bm} 
\usepackage{siunitx} 
\usepackage{xcolor} 
\usepackage[noadjust]{cite} 
\usepackage{threeparttable} 
\usepackage{balance} 

\newcommand{\highlighttext}[1] {#1}

\makesavenoteenv{table}
\makesavenoteenv{table*}
\makesavenoteenv{tabular}

\newcommand{\R}[1] {\textnormal{#1}}

\begin{document}
 
\title{Learning to Detect Open Carry and Concealed Object with 77GHz Radar}

\author{Xiangyu Gao,~\IEEEmembership{Student Member,~IEEE}, Hui Liu,~\IEEEmembership{Fellow,~IEEE}, Sumit Roy,~\IEEEmembership{Fellow,~IEEE}, \\Guanbin Xing,~\IEEEmembership{Member,~IEEE}, Ali Alansari, Youchen Luo
\thanks{X. Gao, H. Liu, S. Roy, G. Xing, A. Alansari are with the Department of Electrical and Computer Engineering, University of Washington, Seattle, WA, 98195, USA. (email: xygao@uw.edu, huiliu@uw.edu, sroy@uw.edu, gxing@uw.edu, afaa97@uw.edu) Y. Luo is with the Paul G. Allen School of Computer Science and Engineering, University of Washington, Seattle, WA, 98195, USA. (email: yluo6@uw.edu)}}

\maketitle

\begin{abstract}
Detecting harmful carried objects plays a key role in intelligent surveillance systems and has widespread applications, for example, in airport security. In this paper, we focus on the relatively unexplored area of using low-cost \SI{77}{GHz} mmWave radar for the carried objects detection problem. The proposed system is capable of real-time detecting three classes of objects - laptop, phone, and knife - under open carry and concealed cases where objects are hidden with clothes or bags. This capability is achieved by the initial signal processing for localization and generating range-azimuth-elevation image cubes, followed by a deep learning-based prediction network and a multi-shot post-processing module for detecting objects. Extensive experiments for validating the system performance on detecting open carry and concealed objects have been presented with a self-built radar-camera testbed and collected dataset. Additionally, the influence of different input formats, factors, and parameters on system performance is analyzed, providing an intuitive understanding of the system. This system would be the very first baseline for other future works aiming to detect carried objects using \SI{77}{GHz} radar. 
\end{abstract}


\section{Introduction}
\IEEEPARstart{A}{bility} to detect person-borne threat objects remains an ongoing and pressing requirement in many scenarios such as airports, schools, and railway stations. \highlighttext{Nowadays, there has been an increased security threat caused by terrorist groups, hijackers, and people hiding weapons in public areas. Thus, an early recognition technology for detecting concealed weapons and triggering an alarm can be a beneficial tool for surveillance purposes. Indeed there are publicity and security measures to prohibit the carrying of dangerous goods, e.g., millimeter-wave (\textbf{mmWave}) body screening at airports, but the slow and complicated imaging process blocks higher passenger throughput rates \cite{gao2021mimosar} and cannot meet the demand of real-time security. Therefore, it is necessary to carry out non-contact human safety inspection for people who may carry dangerous substances.} 

Various sensors have been used for carried object detection. Recently, surveillance cameras with the ability to automatically detect weapons and raise alarms are developed using state-of-the-art deep learning models \cite{9353483, weapon}. However, cameras cannot deal with object blocking or occlusion problems and also pose privacy concerns \cite{ramp}. To address it, numerous technologies which utilize different parts of the electromagnetic spectrum are considered for detecting open carry and concealed objects on persons \cite{4239032}, e.g., using ultrasound \cite{ultra2002}, mmWave \cite{8536660}, Terahertz \cite{4682606}, infrared \cite{817171}, fusion of visual RGB image and infrared \cite{identf2012}, X-ray \cite{Roomi2012DETECTIONOC}, etc. 

\begin{figure}[!t]
  \includegraphics[width=0.47\textwidth]{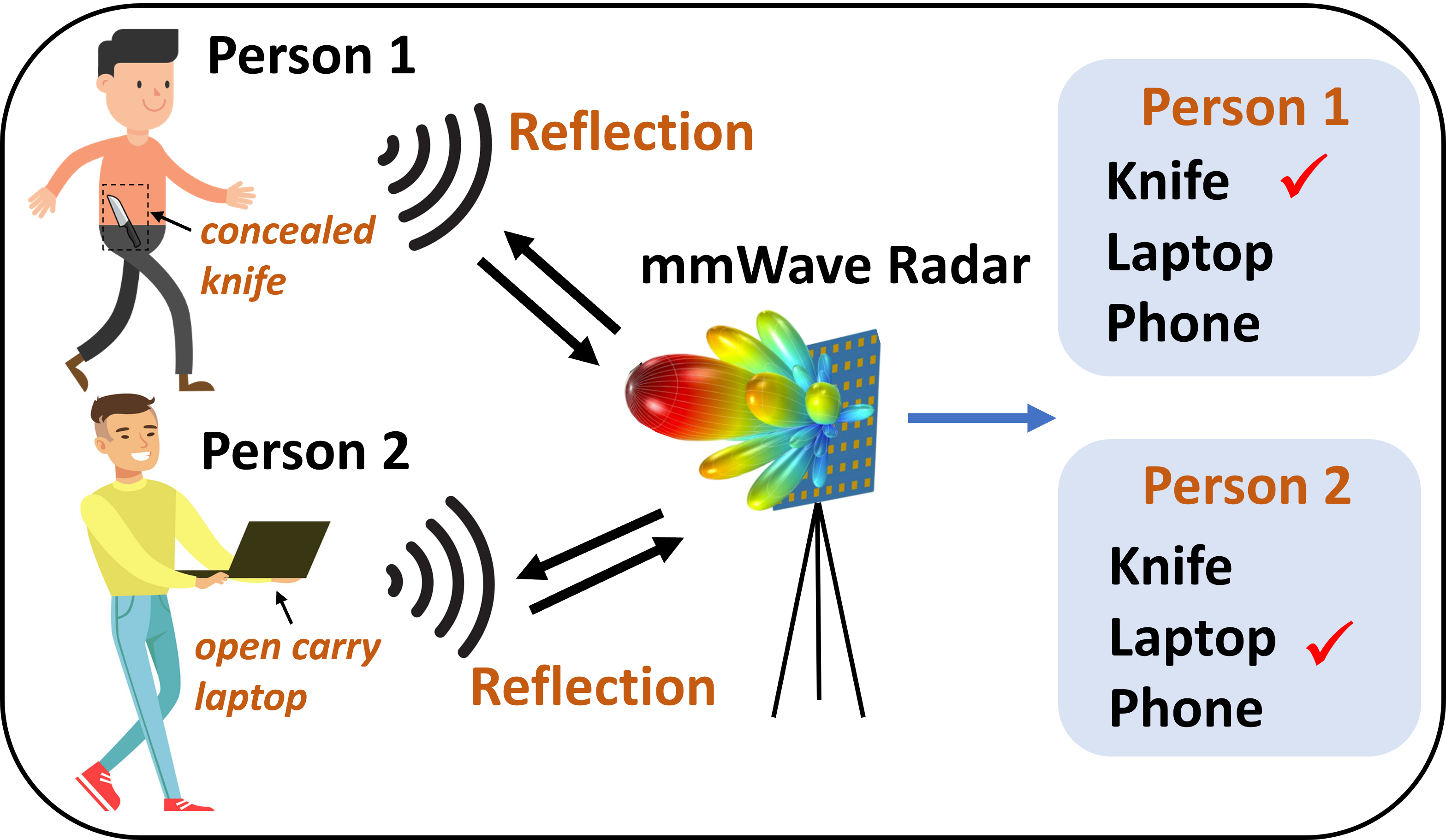}
  \caption{Example usage scenario of carried object detection system: detecting a concealed knife and a open carry laptop on two pedestrians.}
  \label{fig:teaser}
\end{figure}

\highlighttext{The majority of existing weapon object detection algorithms \cite{4682606, 5666180, 11198926, 5530374, identf2012, 8650148} for electromagnetic waves are based on screening (imaging) and contrast (bright spot) detection under the assumption of different reflection coefficients between carried objects and human body \cite{6902403, Essen2006ConcealedWD}. However, a high-resolution imaging result usually takes long processing time or expensive hardware cost. Further, beyond model-based image contrast detection  \cite{identf2012, 5666180, 8650148}, deep learning-based methods were adopted for model-free and accuracy-improved weapon detection on radar imaging, e.g., Faster R-CNN \cite{8628238}, YOLO2 \cite{9269991}. Besides the `detection by image'-based methods, prior works proved to identify a potential moving shooter carrying a concealed rifle by recognizing the unique gait signatures \cite{8519763, 8536660}. It however might not work when someone excels at not showing any physical signs or posture of carrying a weapon or just stands still.}

\highlighttext{In this paper, we apply commercial \SI{77}{GHz} mmWave radar to the carried object detection problem. Specifically, this problem is narrowed down to detecting the existence of three representative objects on pedestrian subjects - laptop, phone, and knife that are common in life and shape-varying - to verify the feasibility. After validation, the proposed system can be easily extended to achieve the detection of dangerous objects, like gun, knife in public places for example airport passageway to provide non-contact human safety inspection. The reason why we use \SI{77}{GHz} mmWave radars is the low price, impressive range resolution (can achieve \SI{4}{cm} with \SI{4}{GHz} bandwidth), fine Doppler velocity discrimination \cite{gao2019experiments}, high angular resolution via signal processing \cite{gao2021mimosar}, and the robust performance under harsh weathers \cite{gao2021perception}. They become more frequently used in autonomous driving for environment imaging \cite{gao2021mimosar, gao2021perception}, semantic object detection \cite{ramp}, occupancy grid mapping \cite{gao2021deform} and in the security industry for person re-identification \cite{9420713} and fall activity detection \cite{6945894}, etc.} 
 
With the \SI{77}{GHz} radar, we proposed a deep learning-based carried object detection (\textbf{COD}) system that takes raw ADC data as input and outputs the predicted existence probabilities for three classes of objects. The COD framework has three main modules: \textit{preprocessing, single-shot prediction network, and multi-shot decision}. The preprocessing module is responsible for detecting targets from raw ADC radar data (I-Q samples post demodulated at the receiver) and cropping small range-azimuth-elevation (\textbf{RAE}) cubes from generated radar imaging based on the detection location. The cropped cube depicts the 3-dimension imaging of a pedestrian carrying objects. Second, the single-shot prediction module is a pyramidal-feature-hierarchy convolutional neural network that takes a single cropped cube as input to make the existence prediction for three classes of objects. Here, we use the combination of preprocessing and prediction (or classification) network, instead of the SSD \cite{Liu_SSD}-like end-to-end neural network for object detection, to reduce the network size or complexity and relieve labeling workload. To further improve system performance, a multi-shot decision module was designed to track the cropped cubes and make final decision via voting from the results of multiple in-track cubes.

For experimenting purposes, a large radar raw data and camera image dataset for a pedestrian subject with various open carry or concealed objects have been collected using the self-built radar-camera testbed. In particular, significant effort was placed in collecting data for situations where cameras are largely ineffective, i.e. objects hidden or covered with clothes. The system performance is analyzed under different scenarios to determine the influence of different input, factors, and parameters on model. The experimental results indicate that \SI{77}{GHz} radar-based COD system performs very well for openly carried objects, and it also works to detect concealed objects in cases with substantial clothing or bag occlusion where camera-based detection does not work. This system would be the very first baseline for other future works aiming to detect carried objects using \SI{77}{GHz} radar. 



In summary, the main novel contributions of this paper are three-fold:
\begin{itemize}
\item A new deep learning-based carried object detection system designed for \SI{77}{GHz} mmWave radar. To the best of our knowledge, we are the first ones applying the RAE imaging results of a commercial automotive radar on this problem in real-world scenes.
\item Extensive experiments for validating the system performance on detecting open carry and concealed objects with self-built testbed and dataset.
\item Analysis of the influence of different input formats, factors, and parameters on system performance, providing an intuitive explanation of the system.
\end{itemize} 

The rest of this paper is organized as follows. The related works and the principle of FMCW MIMO radar are introduced in Section \ref{sec:related_work} and \ref{sec:primer}. The proposed COD system framework is presented in Section \ref{sec:sys}. The system implementation details including the testbed and dataset are described in Section \ref{sec:implment} while the evaluation results are described in Section \ref{sec:res}. We discuss and analyze the system performance in Section \ref{sec:discuss}. Finally, we conclude this paper and propose the future work. 

\section{Related Work \label{sec:related_work}}
\highlighttext{Traditional concealed object detection approaches usually requires the high-resolution imaging from the electromagnetic-wave sensors, which can be achieved by illuminating with large antenna aperture ($\sim$\SI{1}{m}) \cite{4682606, 5666180} or scanning with moving antenna to synthesis aperture \cite{gao2021mimosar, 11198926, 5530374}. With different illumination ways, the imaging can also be divided into two categories: passive sensing \cite{Essen2006ConcealedWD, 5543714, 5666180} and active sensing \cite{gao2019experiments, gao2021perception, 8628238, 8536660, 8519763}. While passive sensing is done with natural illumination (or an incoherent noise source) and takes a long imaging time (a few minutes \cite{Essen2006ConcealedWD}), the active sensing sends a signal via transmitter and then receives the reflected signal from forehand objects, e.g., frequency-modulated continuous-wave (\textbf{FMCW}) radar \cite{gao2019experiments, gao2021perception}, which offers much higher imaging speed than passive sensing but decreased image quality due to the reflection scattering issues \cite{Essen2006ConcealedWD}.}

\highlighttext{The imaging process is followed by the contrast detection to identify weapon contour or edge using various methods like local binary fitting \cite{identf2012}, two-level expectation maximization \cite{5666180}, and Gaussian mixture model \cite{8650148}. Further, deep learning-based methods for example the Faster R-CNN, YOLO2 were adopted in \cite{8628238, 9269991} for model-free and accuracy-improved weapon detection. Other than above contrast detection methods, the gait-based anomaly detection was proposed in \cite{8519763, 8536660} to identify a potential shooter carrying rifle by classifying the micro-Doppler and range-Doppler signatures generated from \SI{77}{GHz} mmWave FMCW radar.}


\begin{figure}[h]
\centering
\includegraphics[width=0.48\textwidth, trim=1 2 1 1,clip]{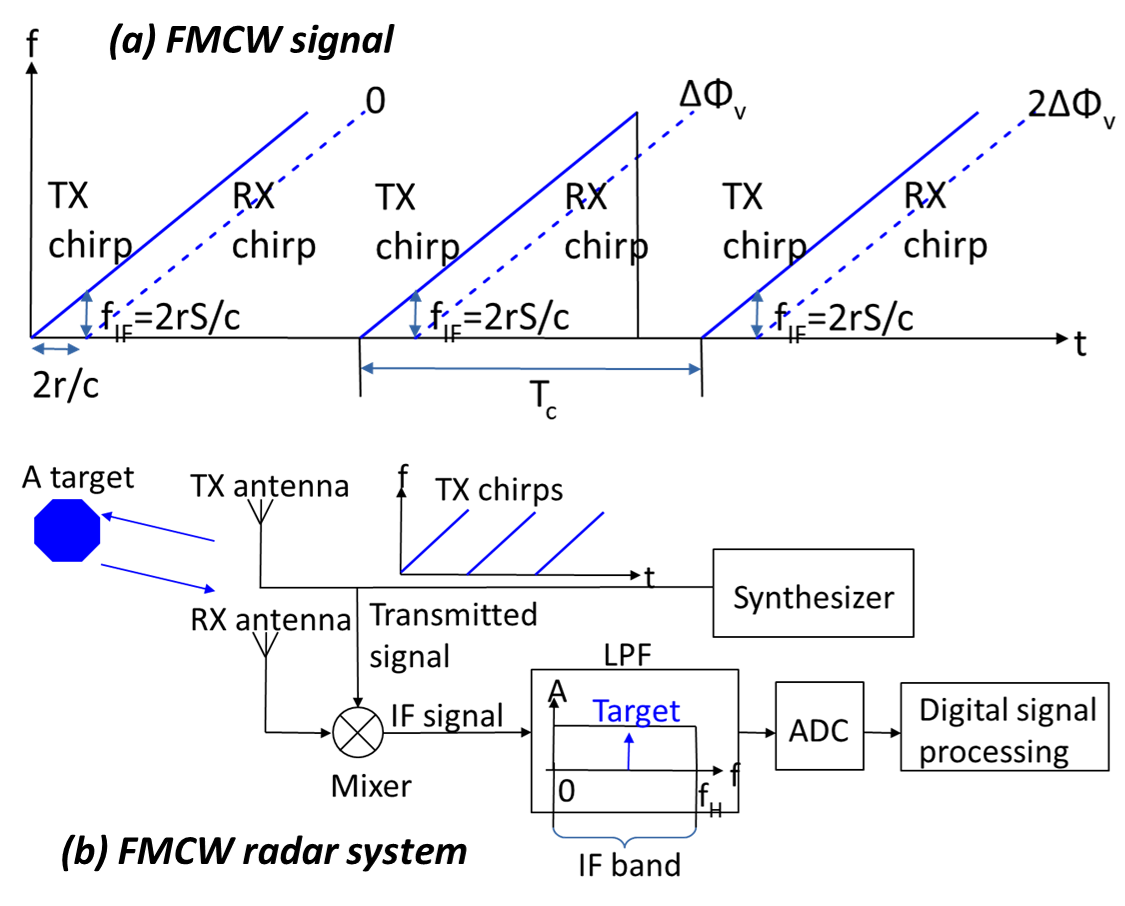}
\caption{\highlighttext{(a) FMCW signal example with 3 TX chirps (blue line) and RX chirps (blue dotted line). The resulted IF signal has frequency $f_{\R{IF}}$ (determined by range $r$) and Doppler phase shift $\Delta \Phi_v$. (b) A FMCW radar system that contains transmitter, receiver, mixer, LPF, ADC, and digital signal processing.}}
\label{fig:fmcw}
\end{figure}

\section{Primer}
\label{sec:primer}
\subsection{FMCW Radar}
FMCW radar transmits a periodic wideband linear frequency-modulated (LFM, also called \textbf{chirps}) signal as shown in Fig.~\ref{fig:fmcw}(a). The transmitted (\textbf{TX}) signal is reflected from targets and received at the radar receiver. FMCW radars can detect targets' range and velocity from the received (\textbf{RX}) signal using the stretch or de-chirping \cite{gao2019experiments} processing structure in Fig.~\ref{fig:fmcw}(b). The mixer at the receiver multiplies the RX signal with the TX signal to produce an intermediate frequency (\textbf{IF}) signal. At the receiver end, the IF signal is passed into an anti-aliasing low-pass filter (LPF) and an analog-to-digital converter (\textbf{ADC}) for the following digital signal processing.
\begin{figure*}[!t]
  \includegraphics[width=1.0\textwidth]{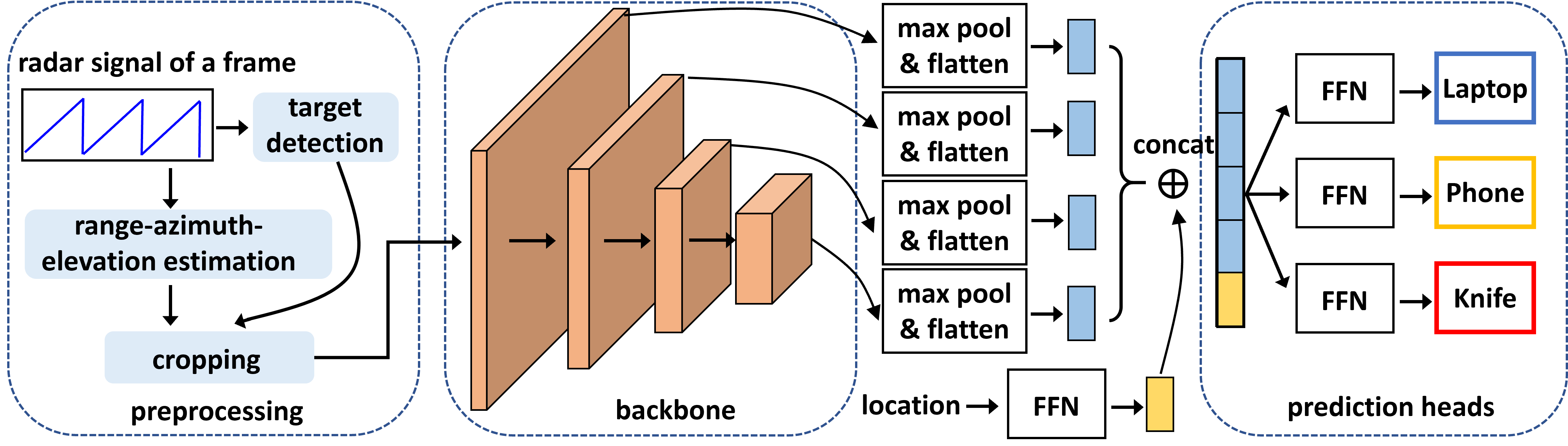}
  \caption{Overview of the proposed carried object detection system with the preprocessing module, and the single-shot prediction network composed of a backbone, feature concatenation, and three prediction heads.}
  \label{fig:system_ss}
\end{figure*}

\textbf{Range estimation}: Since the RX and the TX signal are both LFM signals with constant frequency difference and time delay, the IF signal has a single tone corresponding to the target's distance. For example, the IF frequency for a target at range $r$ is given by $f_\R{IF} = \frac{2r}{c}S$, the multiplication of round-trip delay $\frac{2r}{c}$ and chirp sweeping slope $S$, where $c$ denotes the speed of light. Thus, detecting the frequency of the IF signal can determine the target range. A cost-efficient fast Fourier transform (\textbf{FFT}) is widely adopted here to estimate $f_{\R{IF}}$, and we name it \textbf{Range FFT}.

\textbf{Doppler velocity estimation}:  
Any radial motion $\Delta r$ relative to the radar between consecutive chirps will cause a frequency shift $\Delta f_{\R{IF}} = \frac{2S\Delta r }{c}$ as well as a phase shift $\Delta \phi_v = 2\pi f_{\R{c}}\frac{2\Delta r}{c}=\frac{4\pi v T_{\R{c}}}{\lambda}$ in IF signal \cite{gao2019experiments, iovescu2017fundamentals}, where $f_{\R{c}}$ is the carrier frequency, $v$ is the object velocity, $T_{\R{c}}$ is the chirp period, and $\lambda$ is the wavelength. Compared to the IF frequency shift, the phase shift is more sensitive to the object movement \cite{iovescu2017fundamentals}. Hence, by estimating the phase shift using FFT (named \textbf{Velocity FFT}) across chirps, we can transform the estimated phase to Doppler velocity.

\textbf{Angular estimation}: 
Angle estimation is conducted via processing the signal at a receiver array composed of multiple elements. The return from a target located at far field and angle $\theta$ results in the steering vector \eqref{eq:steering_vec} as the uniform linear array output \cite{526899}:
\begin{equation}
\label{eq:steering_vec}
\resizebox{.91\hsize}{!}{
$\boldsymbol{a}_{\R{ULA}}(\theta)=[1,\ e^{-j2\pi d\sin{\theta}/\lambda},\ \cdots, e^{-j2\pi (N_{\R{Rx}}-1)d\sin{\theta}/\lambda}]^\R{T}$}
\end{equation}
where $d$ denotes the inter-element distance. The embedded phase shift $e^{-j2\pi d\sin{\theta}/\lambda}$ can be extracted by a FFT (named {\textbf{Angle FFT}}) to resolve arrival angles $\theta$ \cite{gao2019experiments}.

In summary, the Range, Velocity and Angle FFT operate on the sample, chirp and receiver dimension of the IF signal $s_{\R{IF}}$ separately, and transform it to the image-like spectrum $S$.
\begin{equation}
\label{eq:3dfft}
\resizebox{.91\hsize}{!}{
$S(r, v, \theta)=\mathcal{F}_{\R{angle}}\mathcal{F}_{\R{velocity}}\mathcal{F}_{\R{range}} \left\{ s_{\R{IF}}(\text{sample, chirp, receiver}) \right\}$}
\end{equation}

\subsection{MIMO \& Virtual Array}
The multiple-input and multiple-output (\textbf{MIMO}) radar is efficient in improving angular resolution by forming a virtual array and increasing valid antenna aperture. This is achieved by sending orthogonal signals on multiple TX antennas, which enables the contribution of each TX signal to be extracted at each RX antenna. Hence, a physical TX array with $N_\R{T}$ elements and RX array with $N_\R{R}$ elements will result in a virtual array with up to $N_\R{T}N_\R{R}$ unique (non-overlapped) virtual elements \cite{Wang2012VirtualAA}. The virtual array is located at the spatial convolution of TX antennas and RX antennas, i.e., convolution produces a set of virtual element locations that are the summed locations of each TX and RX pair. To reduce array cost (fewer physical antenna elements), non-uniform arrays spanning large apertures have been proposed, e.g., the minimum redundancy array \cite{gao2021perception, mra}. 

When performing angular estimation on a MIMO virtual array, the motion-induced phase errors (i.e., for non-stationary targets) should be compensated on virtual elements before performing Angle FFT \cite{gao2021mimosar}. The motion-included phase difference has to be considered under time-division multiplexing (\textbf{TDM}) scheme because of the switching time between the transmitters. The correction of phase error can be done via compensation of half the estimated Doppler phase shift ($\Delta \phi_{\R{v}}/2$) obtained from the Velocity FFT results \cite{8052088}.

\section{System Design}
\label{sec:sys}
The proposed carried object detection (COD) system has three main modules - preprocessing, single-shot prediction network, and multi-shot decision. The preprocessing module is responsible for detecting targets from the input ADC radar data and cropping small range-azimuth-elevation cubes from the generated radar image for detected targets. The single-shot prediction network takes a single cropped cube as input to make the existence prediction for three classes of objects: laptop, phone, and knife. To further improve system performance, the multi-shot decision module tracks the cropped cubes and makes the final decision based on multiple in-track cubes.

\begin{figure*}[t]
\centering
\includegraphics[width=\textwidth,clip]{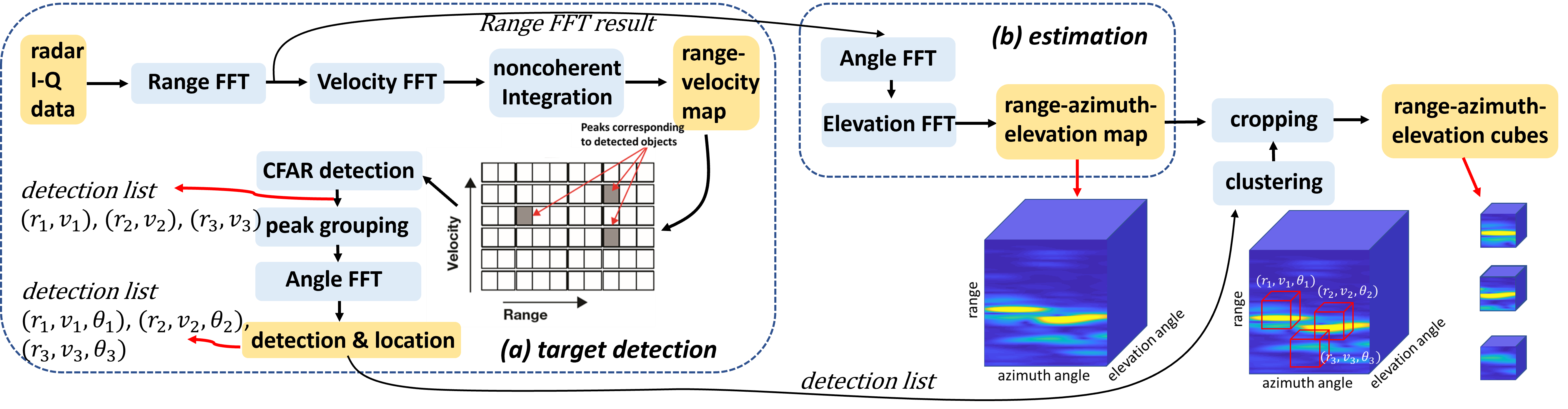}
\caption{Signal preprocessing block diagram where the raw radar I-Q data is processed to obtain the cropped range-azimuth-elevation cubes. The blue blocks represent the operation and yellow blocks represent the input or output.}
\label{fig:prepro}
\end{figure*}

\subsection{Preprocessing}
\textbf{Target detection}: The target detection and localization are achieved by processing FMCW radar I-Q samples and applying constant false alarm rate (\textbf{CFAR}) detection algorithm, as shown in Fig.~\ref{fig:prepro}(a). First, the Range and Velocity FFTs are performed on I-Q data in a frame to obtain the range-velocity (\textbf{RV}) map for initial target detection. The RV maps from all receivers are integrated non-coherently (i.e., sum the magnitude RV maps) to increase the signal-to-noise ratio of the resulting RV map. Post summing, the 1D cell-averaging CFAR \cite{piofmd} algorithm is applied along the Doppler dimension and range dimension separately to detect targets or peaks and obtain their 2D localization (range, velocity). During the CFAR detection process, each cell or bin is evaluated for the presence or absence of a target using a threshold that adapts itself according to the noise power estimated within a sliding window.

Thereafter, peak grouping for all CFAR detections is done by checking if each detection has a greater amplitude than its neighbored detections. For example, there is detection \#2 that lies within the $3\times3$ range-velocity kernel centered at detection \#1, then detection \#2 will be discarded if it has a smaller amplitude \cite{gao2021mimosar}. Peak grouping is intended to simply cluster the very close peaks or detections. Subsequently, we estimate azimuth angles for the remaining detections and obtain their final localization in range, velocity, and azimuth angle. This is done by calculating the Angle FFT for each detected target across the RV maps of all receivers (i.e., complete virtual array in TDM-MIMO case). Following \cite{gao2021mimosar, 8052088}, we compensate the motion-induced phase error for TDM-MIMO using the estimated Doppler velocity before Angle FFT.

\textbf{Range-azimuth-elevation estimation}: To get a 3D view of pedestrian subjects and their carried object, the range-azimuth-elevation estimation is implemented for radar imaging. The imaging result for each frame is a 3D spectrum with the range dimension, azimuth angle dimension, and elevation angle dimension. As the range estimation has already been done in the target detection part, the remaining azimuth-elevation estimation processing is continued on the Range FFT output. 
That is, we perform the Angle FFT for azimuth angle estimation and another FFT for elevation estimation on it. The first Angle FFT works across all horizontal elements of the 2D receiver array we have in Fig.~\ref{fig:mimo}, while the second FFT works across all vertical elements. Details of this workflow are illustrated in Fig.~\ref{fig:prepro}(b).

\textbf{Clustering and Cropping}: To reduce the size of the input to the network, we crop small cubes from the generated 3D range-azimuth-elevation map based on the location of detections, and only input the cropped cubes to the following prediction network. To decrease the total number of cropped cubes, we apply clustering for detections with their localization before the cropping operation. A parameter $\boldsymbol{\varepsilon}=[\varepsilon_r, \varepsilon_v, \varepsilon_a]$ is defined to specify how close points should be to each other on range, velocity, and azimuth angle dimension to be considered a part of a cluster. It means that if the distance between two detections is lower or equal to this value $\boldsymbol{\varepsilon}$, these detections are considered neighbors. The center location of the resulting clusters are taken as new detections, and for each of it we crop a cube centered on the specified range, azimuth angle, and zero elevation angle. It is assumed that each cropped cube contains a pedestrian subject with carried object when an appropriate clustering threshold $\boldsymbol{\varepsilon}$ is used. The cube size is set to $24 \times 24 \times 10$ to cover most of the region of human body even at close range. 

\subsection{Single-shot Prediction Network \label{sec:sspn}}

\begin{figure}
\centering
\includegraphics[width=0.49\textwidth,trim=2 3 3 2,clip]{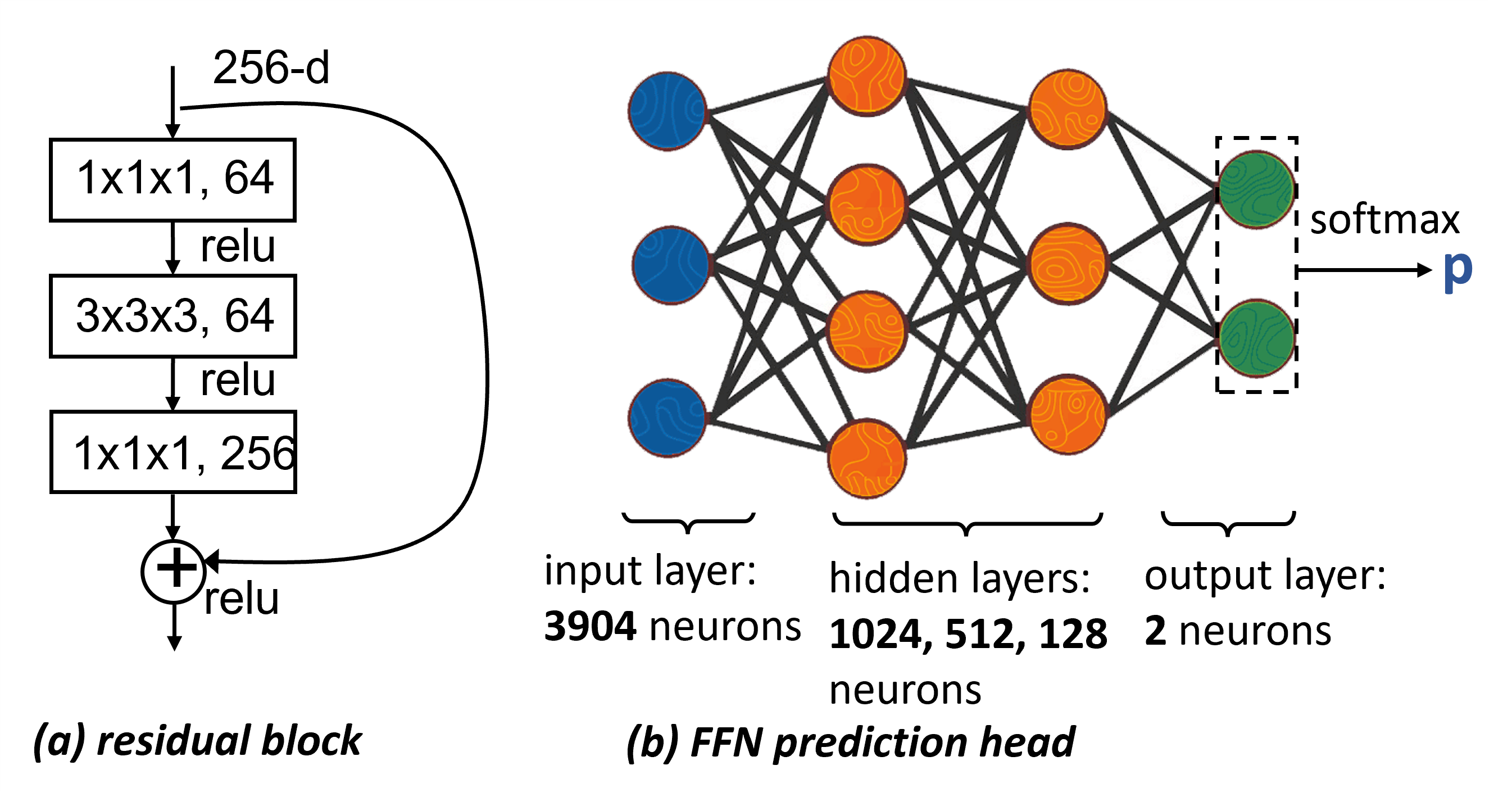}
\caption{\highlighttext{(a) Example of a residual building block for ResNet-50. It contains a stack of 3 layers and an identify shortcut. (b) Illustration of a 5-layer FFN prediction head.}}
\label{fig:layers}
\end{figure}

\textbf{Backbone}: The backbone is a deep residual pyramid neural network that takes a single cropped range-azimuth-elevation cube as input for feature extraction. The backbone is modified from ResNet-50 \cite{7780459} and it has 49 \textit{3D convolutional layers}. The convolutional layers mostly have $3\times3\times3$ filters and $1\times1\times1$ filters and they are mainly divided into 5 parts (conv1$\underline{\enskip}$x, conv2$\underline{\enskip}$x, conv3$\underline{\enskip}$x, conv4$\underline{\enskip}$x, and conv5$\underline{\enskip}$x) shown in Table.~\ref{tab:layers}. We perform downsampling at the end of the last four parts directly by convolutional layers with a stride of 2. Except the first part, each has several three-layer bottleneck blocks for performing \textit{residual function by shortcut connection} \cite{7780459}. One residual block example is presented in Fig.~\ref{fig:layers}(a).

\textbf{Feature concatenation}: The convolutional backbone computes a feature hierarchy layer by layer, and with downsampling, the feature hierarchy has an inherent multi-scale and semantic gaps \cite{fpn}. \highlighttext{Particularly in our case the local features from lower layers would be beneficial for carried object detection since the object is occupied part of the input image of pedestrian subject.} Following this idea, we reuse the multi-scale feature maps from different layers computed in the forward pass that comes free of cost, as illustrated in Fig.~\ref{fig:system_ss}. The multi-scale feature maps go through \textit{max-pooling layers} to  extract sharp patterns \cite{7780459} and reformat output to same size $1\times1\times1$. Then the reformatted multi-scale feature maps are flattened and concatenated together to form the multi-scale features ($1\times 3840$ size). Besides, location features of the cropped cube ($1\times 64$ size) is extracted by passing the center location (i.e., range and azimuth angle) to a 3-layer feed-forward neural network (\textbf{FFN}). The obtained location features and multi-scale features are then concatenated together as the final extracted features ($1\times 3904$ size). 

\textbf{Prediction heads}:
The concatenated features are input to three same prediction heads to infer if there exists a carried laptop, phone, or knife for the input cube. Each prediction head is a 5-layer FFN that makes the existence prediction for a class. For example, if the third FFN is in charge of observing knives and it will output the probability of carrying a knife $p$. We then can simply check if $p> p_\R{thr}$ to make the single-shot prediction, where $p_\R{thr}$ is a probability threshold. The FFN used here has 1 input layer, 3 hidden layers, and 1 output layer as shown in Fig.~\ref{fig:layers}(b). The last layer has 2 outputs $(o_1, o_2)$, which are transformed to the prediction probability $p$ using the \textit{softmax function} as follows.
\begin{equation}
\label{eq:softmax}
p=\frac{\exp(o_{1})}{\exp(o_{1}) + \exp(o_{2})}
\end{equation}

It is worth to be noted here that we adopt 3 independent binary prediction heads instead of using one 3-class prediction head. The reason behind it is that the 3-class prediction head with \textit{softmax} doesn't allow the coexistence of more than one object which is pretty common in real-life scenarios.

\textbf{Loss function}:
The loss function for proposed network \eqref{eq:loss} is the weighted sum of Focal Losses \cite{focal} from all three prediction heads. Focal Loss is adopted here to address class imbalance during training, i.e., for each binary prediction head, the number of corresponding object (e.g., laptop) is naturally less than the total number of non-objects (e.g., phone and knife). Focal loss applies a modulating term $(1-p)^\alpha$ to the cross-entropy loss in order to focus learning on hard negative examples \cite{focal}.
\begin{equation}
\label{eq:loss}
\begin{aligned}
&\qquad \qquad \text{Loss} = \text{FL}_\R{laptop} + \text{FL}_\R{phone} + \text{FL}_\R{knife}\\
&\text{FL}(p) = - w_1 y (1-p)^\alpha \log p - w_2 (1-y) p^\alpha \log (1-p)
\end{aligned}
\end{equation}

\noindent where $y=0~\text{or}~1$ is the ground truth and $p$ is the predicted existence probability for a certain class of objects. $\alpha$ is a tunable focusing parameter, $w_1$ and $w_2$ are weight-balance parameters.

\subsection{Multi-shot Decision}
We have mentioned the single-shot prediction network in Section~\ref{sec:sspn} with one cropped cube as input. To further improve detection performance by introducing different observation perspectives from multiple frames, a simple multi-shot decision scheme is proposed - to track the cropped cubes and make voting decision based on multiple in-track cubes.

\begin{figure}
\centering
\includegraphics[width=0.45\textwidth,clip]{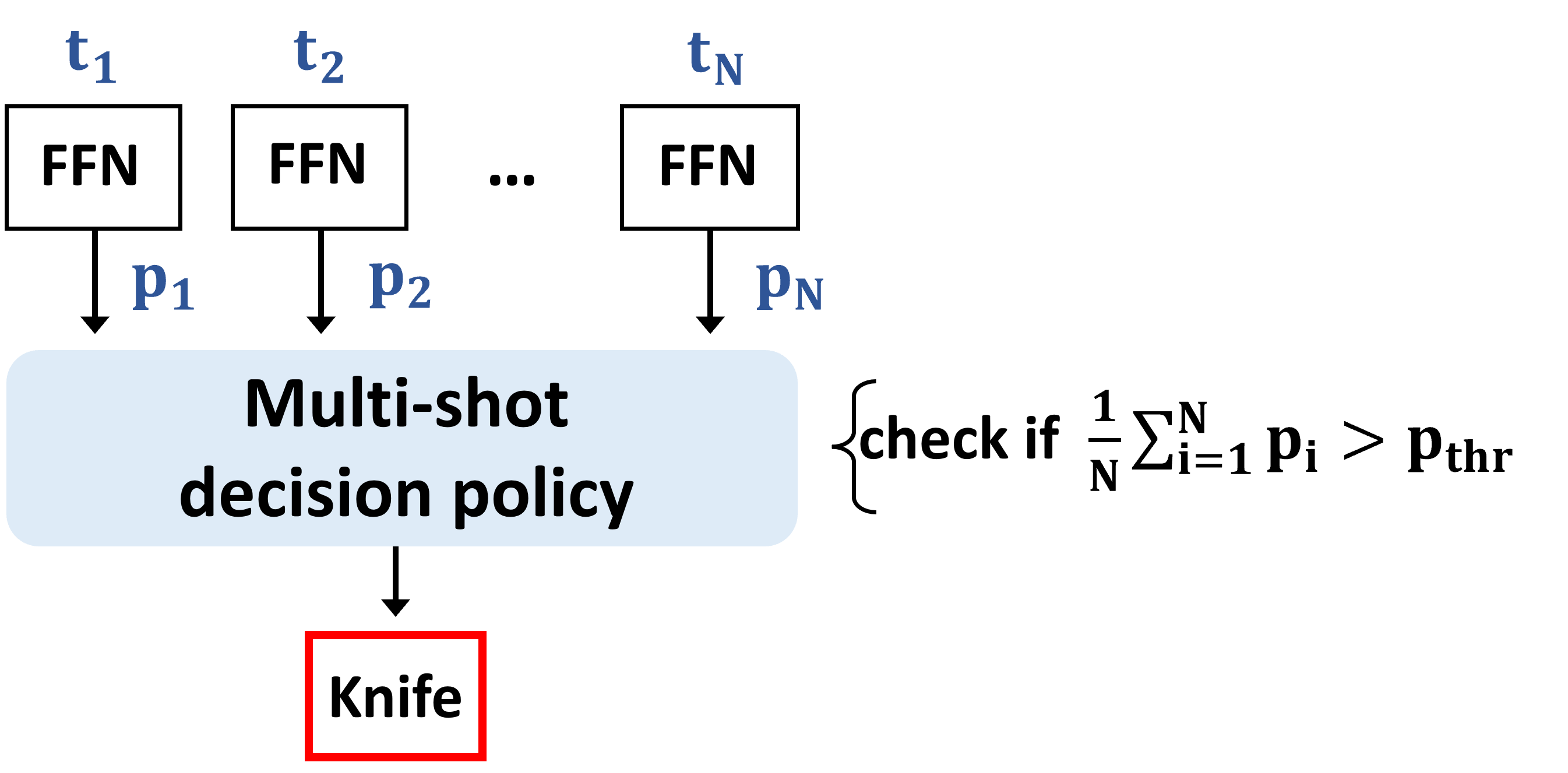}
\caption{\highlighttext{Explanation of multi-shot decision policy for detecting knife. The prediction probabilities $p_1,\, p_2,\, \dots,\, p_N$ for $N$ timestamps are averaged and then compared with the threshold $p_{\R{thr}}$.}} 
\label{fig:multishot}
\end{figure}

\begin{figure*}[!t]
  \includegraphics[width=0.9\textwidth]{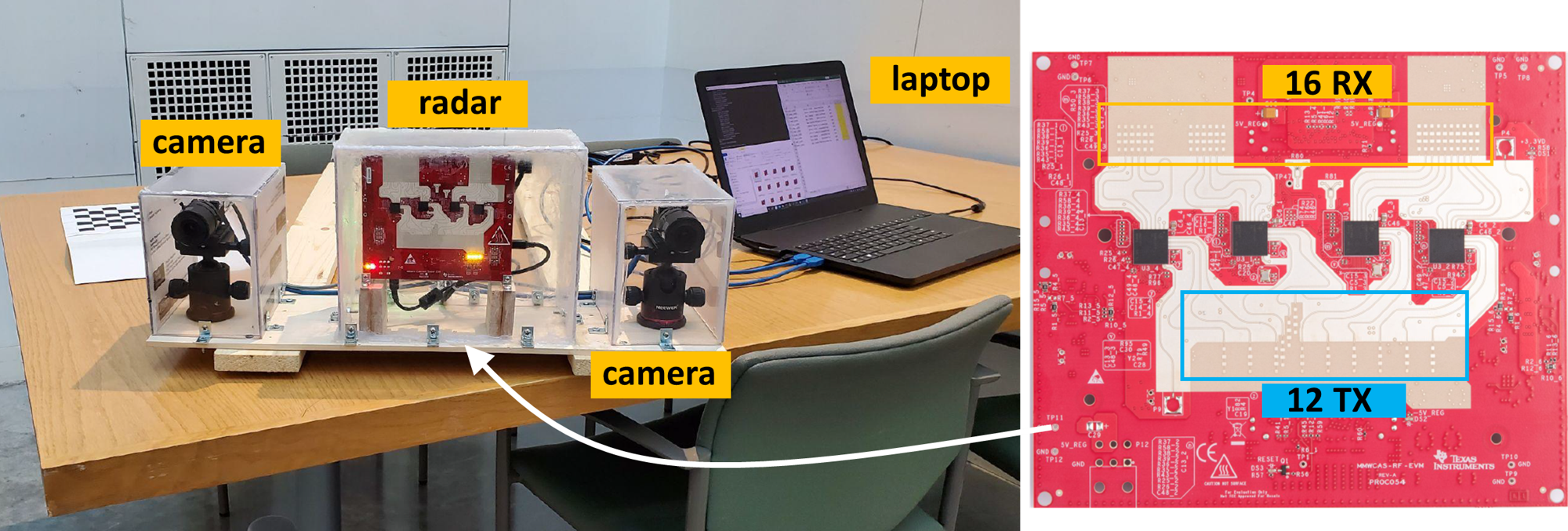}
  \caption{\highlighttext{Left: Experimental radar-camera testbed that consists of a TIDEP-01012 cascaded-chip mmWav radar from Texes Instrument and two cameras from FLIR. The testbed is configured and controlled by a laptop for data collection. Right: the front view of TIDEP-01012 radar where 4 chips (black squares, 12 TX (blue box), and 16 RX (orange box) are incorporated.}}
  \label{fig:testbed}
\end{figure*}

\begin{figure}[t]
  \includegraphics[width=0.47\textwidth]{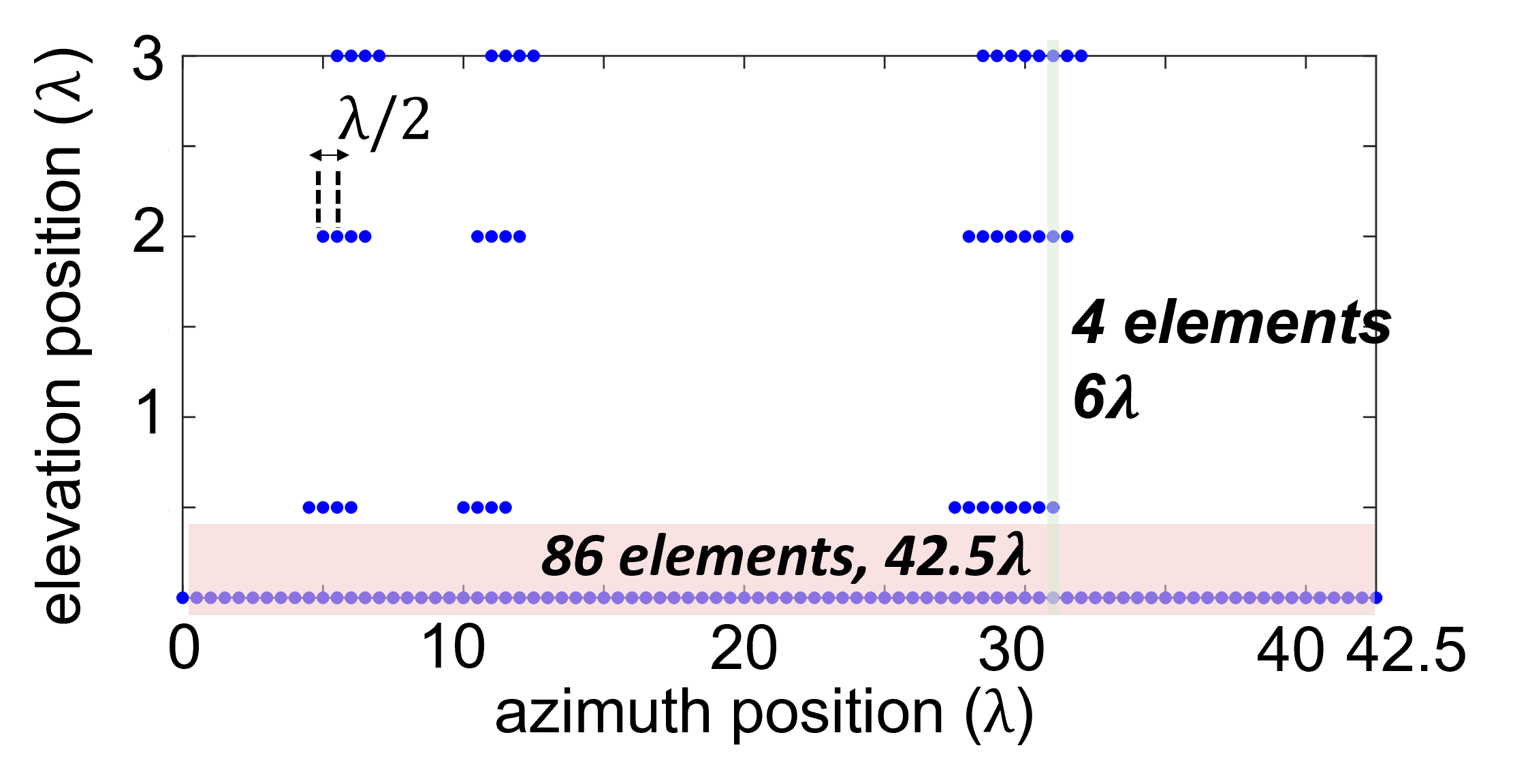}
  \caption{\highlighttext{2D virtual array formed by 12 TX and 16 RX with TDM-MIMO configuration. It spans the azimuth dimension in $42.5\lambda$ and the elevation dimension in $3\lambda$, with 192 virtual elements in total. Except the bottom-row 86-element linear array, this virtual array is mostly sparse elsewhere.}}
  \label{fig:mimo}
\end{figure}

\textbf{Tracking cropped cubes}: A Kalman filter operating on subsequent frames is applied to obtain a reliable estimation of the true subject’s state (i.e., its location). The association of the cropped cubes detected in the current time frame with the right user trajectories is performed using the Hungarian algorithm.

\textbf{Decision policy}: Assume we have $N$ in-track cropped cubes that belong to same  pedestrian carrying an object and input them separately to the single-shot prediction network to obtain $N$ independent prediction probability $p_1, \ p_2, \ \dots, \ p_N$. How to make the final decision based on them to achieve more coherent prediction results and therefore better performance?

The proposed method is a simple voting policy that measures average probability over $N$ predictions and checks if averaged $p$ is greater than probability threshold $p_\R{thr}$, as shown in \eqref{eq:vote}. The threshold is tunable for satisfying the system requirement on false alarm and sensitivity.
\begin{equation}
\label{eq:vote}
\frac{1}{N}\sum_{i=1}^N p_i > p_\R{thr}    
\end{equation}

\section{Implementation}
\label{sec:implment}
\subsection{Testbed}

The experiment test-bed (Fig.~\ref{fig:testbed} left) was assembled with a TIDEP-01012 \SI{77}{GHz} mmWave radar \cite{ti_casd} and binocular FLIR cameras. The binocular cameras and radar are connected to the same laptop which uses the timestamp to keep inter-sensor synchronization. The synchronization between the two cameras is achieved by joining them together with an additional cable and using the same trigger clock. The radar data collection pipeline is implemented by combining MATLAB scripts and TI software development kits (SDK) while the camera pipeline is implemented by Python scripts and FLIR SDK. \highlighttext{Note that collected camera images are not incorporated into the system processing chain and only used for providing the visualization for experiment scenarios.}

The adopted mmWave radar is a 4-chip cascaded evaluation board with 12 TX antennas and 16 RX antennas (Fig.~\ref{fig:testbed} right). With time-division multiplexing (TDM) on TXs, it can form a large 2D-MIMO virtual array (Fig.~\ref{fig:mimo}) with 192 elements via the spatial convolution of all TX and RX, resulting in fine azimuth resolution (\SI{1.35}{\degree}) and additional elevation resolution (\SI{19}{\degree}). The configuration of this radar is presented in Table.~\ref{tab:sys_param}. Based on those parameters and the calculation equations in \cite{gao2019experiments}, we can give out the capability of this radar in terms of range resolution ($\frac{c}{2B}=\SI{0.06}{m}$), max detectable range ($\frac{f_{\R{s}} c}{2S} = \SI{15}{m}$), Doppler velocity resolution ($\frac{\lambda}{2N_{\R{c}} T_{\R{c}}}=\SI{0.072}{m/s}$), and max operating velocity ($ \frac{\lambda}{4T_{\R{c}}} = \SI{1.80}{m/s}$). 

\begin{table}[t]
\centering
\begin{threeparttable}
\caption{Configuration for adopted mmWave radar}
\begin{tabular}{ll}
\toprule  
 \textbf{Configuration} & \textbf{Value}\\
\midrule  
Frequency ($f_{\R{c}}$) & \SI{77}{GHz}\\[0.75ex]
 
Sweeping Bandwidth ($B$) & \SI{2.5}{GHz}\\[0.75ex]

Sweep slope ($S$) & \SI{79}{MHz \per\micro\second} \\[0.75ex]

Sampling frequency ($f_{\R{s}}$) & \SI{8}{Msps}\\[0.75ex]

Num of chirps in one frame ($N_{\R{c}}$) & $50$ \\[0.75ex]

Num of samples of one chirp ($N_{\R{s}}$) & $256$ \\[0.75ex]

Duration of chirp~\protect \footnotemark~and frame ($T_{\R{c}}$, $T_{\R{f}}$) & \SI{540}{\micro\second}, \SI{1/30}{s} \\
\bottomrule 
\label{tab:sys_param}
\end{tabular}
\begin{tablenotes}
\item [1] $T_\R{c}$ is equal to single chirp interval times number of TX antennas, i.e., $T_\R{c}=\SI{45}{us} \times 12=\SI{540}{us}$.
\end{tablenotes}
\end{threeparttable}
\end{table}

\subsection{Data Collection and Dataset}
\begin{figure}
\centering
\includegraphics[width=0.45\textwidth,clip]{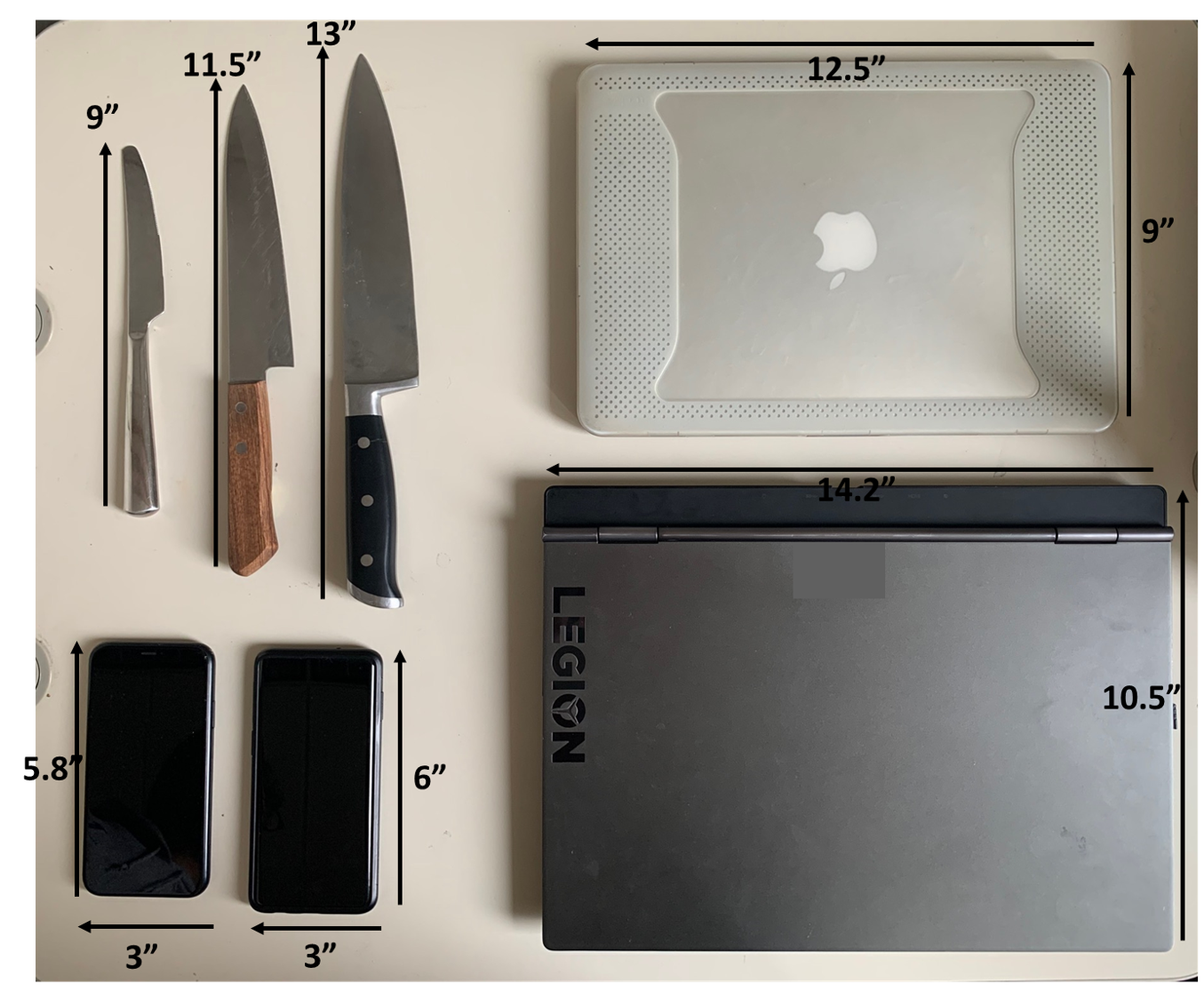}
\caption{A selection of used objects and their sizes, which includes 2 cutting knives, 1 butter knife, 2 regular phones and 2 laptops.}
\label{fig:object}
\end{figure}

\highlighttext{Four main object groups were used during the data collection process: \textit{phones, laptops, knives (include metallic butter knives and cutting knives), and others (e.g., keys, no object)}. Phones, laptops, and keys were selected as they are common objects carried by many people in their daily lives. Since the purpose of the system is safety and security and we were not able to perform data collection with other dangerous objects (e.g. firearms), knives were used as the dangerous item to be detected. The objects that were used varied in weight, size, and shape, as shown in Fig.~\ref{fig:object}. Different laptops, phones, and keys were used throughout the data collection process to increase the variability of the data set. Three subjects were involved in the data collection process, which was done in the building lobby and laboratory room with different device placement locations every time. A single data collection run consisted of a subject holding one of the four object groups listed above, and the subject would walk at a normal pace on a random path for 10 seconds in front of the testbed while either concealing or openly carrying the objects. The testbed would capture 300 frames of camera images and radar raw ADC data at 30 frames per second. To add variability to the data, the walking pattern of subjects was always randomized. Additionally, the location of where the objects were concealed or how the objects were openly carried was always changed.} 

\begin{figure}
\centering
\includegraphics[width=0.5\textwidth,clip]{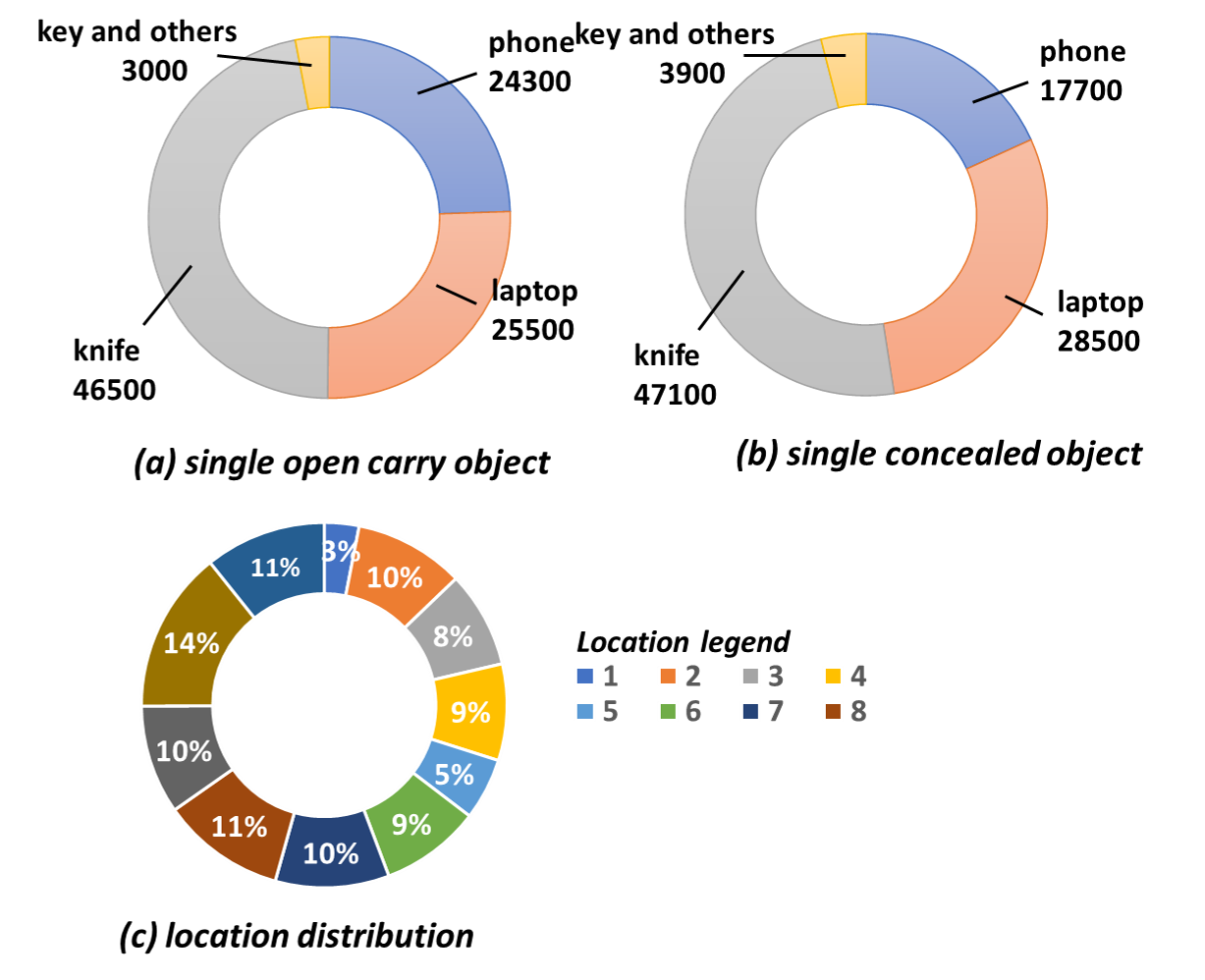}
\caption{Dataset distribution: (a) Data distribution of single open carry object scenario; (b) Data distribution of single concealed object scenario; (c) Data distribution of 8 experiment places.}
\label{fig:dataset}
\end{figure}

The data consisted of \textit{single object being openly carried} and \textit{single object being concealed}. The subjects performed the data runs with different clothing types - low, medium, and heavy - which corresponds to the thickness of the clothing. For example, a t-shirt would be considered low, while a jacket on top of another layer would be heavy. A total of 196500 frames (\highlighttext{lasting 1.82 hours}) were collected for a subject with single object, 99300 of those were open carry and 97200 were concealed. The detailed class distribution and location distribution for the collected dataset are described in Fig.~\ref{fig:dataset}. A sample dataset would be made publicly available to encourage future works \footnote{Please contact us if you are interested in the dataset. (\url{xygao@uw.edu})} 



\subsection{Preprocessing}
The preprocessing was conducted using MATLAB R2020b on a computer with Intel i7-7700K CPU to detect potential targets, and crop the range-azimuth-elevation cubes from the generated 3D radar imaging. During preprocessing, the used hyper-parameters are summarized here: probability of false alarm in CFAR ($1 \times 10^{-4}$), Range FFT points (256), Velocity FFT points (64), Angle FFT points for azimuth (86), Angle FFT points for elevation (16), clustering threshold ($[\varepsilon_r, \varepsilon_v, \varepsilon_a] = [10, 8, 8]$), cropped cube size along range, azimuth and elevation (24, 24, 10), and cube amplitude normalization value ($1 \times 10^5$). The cropped and normalized cubes are stored at the local disk for following training and testing usage. Moreover, for the in-track data used in multi-shot prediction, the Kalman filter and Hungarian algorithm were used to track the cubes from multiple frames and we saved the tracking association results to local disk as well.

\subsection{Training and Testing}
\label{sec:imp-train}
\highlighttext{The single-shot prediction network was implemented using PyTorch and Python libraries, and the training and testing were conducted on a computer with a TITAN RTX GPU.} Particularly, to address the problem we found that the training loss is easily stuck at some points with a large training set initially, a two-step training strategy was used here. That is, a small subset of the training data is used for training the network first to get a pre-trained model, and the pre-trained model is used as initialization for the second training with complete training set. In our experiments, the first training step starts with learning rate $4\times10^{-4}$ and stops when the training accuracy approaches \SI{90}{\percent}, while the second training step starts with smaller learning rate $1\times10^{-4}$. Besides, a batch size of $32$, SGD optimizer, and the learning rate decayed by half every 10 epochs are used in both two steps. 

We trained and tested model on the open carry dataset and concealed dataset separately. For parameters in loss function, we used fixed focusing value $\alpha=2$ but different balance weights $[w_{1,\text{laptop}}, w_{2,\text{laptop}}, w_{1,\text{phone}}, w_{2,\text{phone}}, w_{1,\text{knife}}, w_{2,\text{knife}}]$ for two dataset. For open carry and concealed data, the fine-tuned weights are as follows $[1, 1, 20, 1, 1, 1]$, $[2, 1, 20, 1, 1, 1]$. In addition, we note that the training RAE cubes are generated from a random chirp of frame, aiming to expand training set without any cost. While for testing cubes, we average the imaging results of all chirps within frame to reduce noise and thus improve their signal-to-noise ration. 

\begin{table*}[!ht]
\centering
\caption{Evaluation results for open carry and concealed object using COD-single and COD-multi method.}
\resizebox{\textwidth}{!}{
\setlength\tabcolsep{3pt} 
\begin{tabular}{lcccccccccc}
\toprule
\textbf{Object} & \textbf{Method} & \multicolumn{4}{c}{\textbf{Metric}} & \textbf{Method} & \multicolumn{4}{c}{\textbf{Metric}}\\
\cmidrule(lr){3-6} \cmidrule(lr){8-11}
 & COD-single & precision$\uparrow$ & recall$\uparrow$ (missing$\downarrow$) & false alarm$\downarrow$ & F1$\uparrow$ & COD-multi & precision & recall (missing) & false alarm & F1 \\
\textit{open carry:} \\
(a) laptop & & 0.6146 & 0.858 (0.142) & 0.1948 & 0.716 & & 0.6637 & 0.9132 (0.0868) & 0.1681 & 0.7684 \\
(b) phone & & 0.5059 & 0.7407 (0.2593) & 0.2446 & 0.5995 & & 0.5618 & 0.78 (0.22) & 0.2065 & 0.6511 \\
(c) knife & & 0.7036 & 0.7396 (0.2604) & 0.2769 & 0.7211 & & 0.7645 & 0.809 (0.191) & 0.2217 & 0.7861 \\
(d) average & & 0.6081 & 0.7794 (0.2206) & 0.2388 & 0.6789 & & 0.6633 & 0.8341 (0.1659) & 0.1988 & 0.7352 \\ \midrule

\textit{concealed:} \\
(a) laptop & &  0.4610 & 0.773 (0.277) & 0.3749 & 0.5628  & &  0.5305 & 0.8114 (0.1886) & 0.3191 & 0.6411 \\
(b) phone & &  0.3256 & 0.6915 (0.3085) & 0.3087 & 0.442 & &  0.3743 & 0.7613 (0.2387) & 0.2761 & 0.5005 \\
(c) knife & &  0.5918 & 0.6591 (0.3409) & 0.4462 & 0.6235 & &  0.6256 & 0.7119 (0.2881) & 0.4201 &  0.6657 \\
(d) average & &  0.4595 & 0.6912 (0.3088) & 0.3766 & 0.5428 & &  0.5101 & 0.7615 (0.2385) & 0.3384 & 0.6024 \\
\bottomrule
\end{tabular}}
\label{tab:res_single}
\end{table*}

\begin{figure*}
\centering
\includegraphics[width=1.0\textwidth,clip]{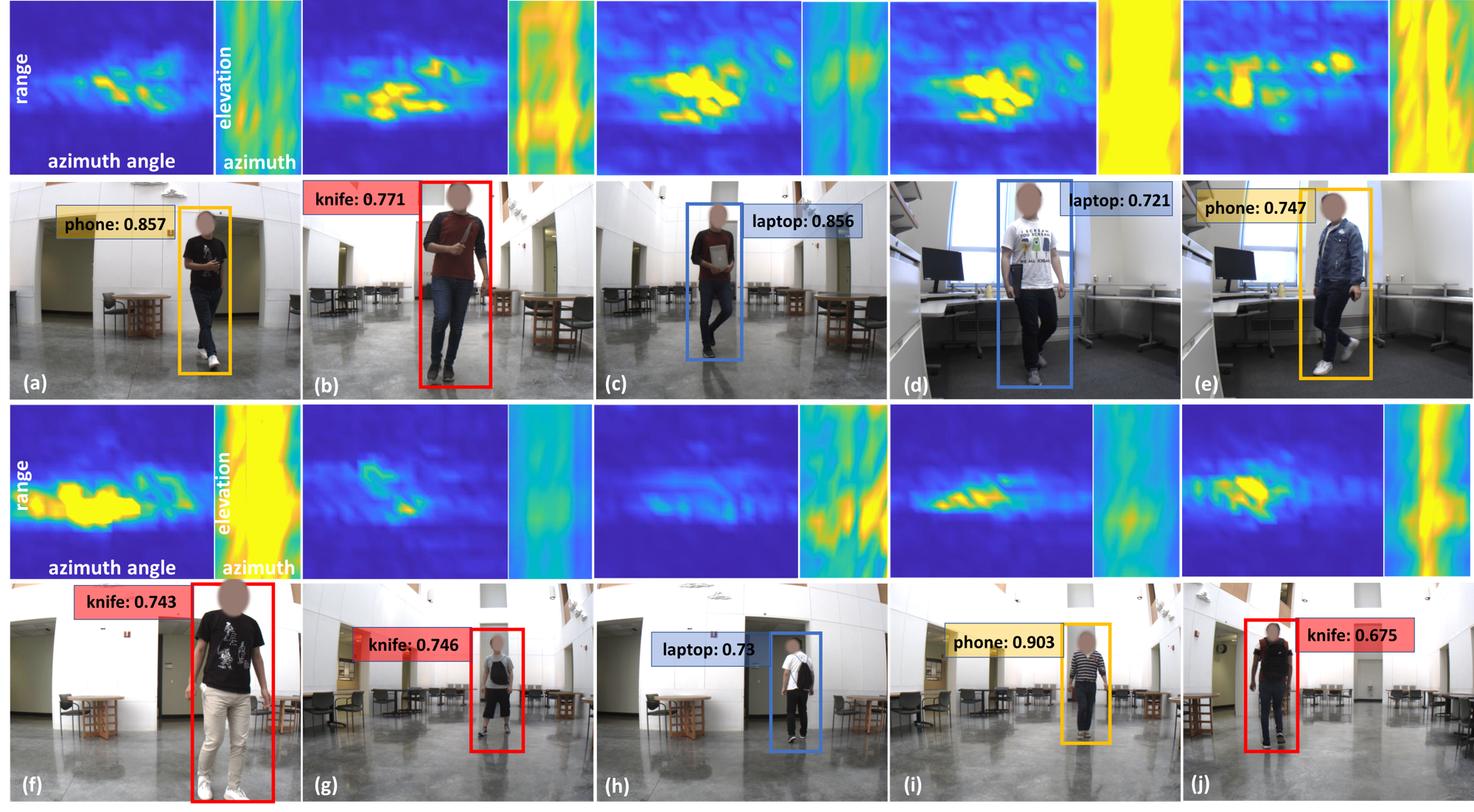}
\caption{Test samples with input cube visualization and detection results. Row 1, 3 show the visualization of the input cropped cube in range-azimuth angle profile and elevation-centered azimuth angle profile. Row 2, 4 show the detection results on camera images with bounding box and detection probability. Note that the bounding box is plotted manually by projecting detection results and ignoring all with probability less than 0.5. First two rows are all open carry objects while last two rows are all concealed ones: (a) subject carrying phone in hand, (b) subject carrying knife in hand, (c) subject carrying laptop in front of body, (d) subject carrying laptop by their side, (e) subject carrying phone by their side, (f) subject carrying knife inside a pocket, (g) subject carrying knife inside a backpack, (h) subject carrying a laptop in a backpack, (i) subject carrying phone inside a pocket, (j) subject carrying knife in a backpack.}
\label{fig:qual_res}
\end{figure*}

\section{Evaluation}
\label{sec:res}
\subsection{Metrics}
\label{sec:res_metric}
Five metrics are adopted for evaluating the effectiveness of our system: \textit{precision}, \textit{recall}, \textit{false alarm (rate)}, \textit{missing (rate)}, and \textit{F1}. They are defined in \eqref{eq:metrics} using true positive (\textbf{TP}), true negative (\textbf{TN}), false positive (\textbf{FP}), and false negative (\textbf{FN}) detections. Specifically, precision means the proportion of positive identifications is actually correct, recall means the proportion of actual positives was identified correctly, and F1 is a measure that combines precision and recall. False alarm represents the probability of falsely detecting something when it does not exist, while missing rate represents the probability of falsely ignoring something when it actually exists.
\begin{equation}
\label{eq:metrics}
\begin{aligned}
& \resizebox{.91\hsize}{!}{$\text{precision}=\frac{\text{TP}}{\text{TP}+\text{FP}},\ \text{recall}=\frac{\text{TP}}{\text{TP}+\text{FN}},\ \text{missing}=1-\text{recall}$} \\
&\resizebox{.7\hsize}{!}{$\text{false alarm}=\frac{\text{FP}}{\text{FP}+\text{TN}},\ \text{F1}=2 \cdot \frac{\text { precision } \cdot \text { recall }}{\text { precision }+\text { recall }}$}
\end{aligned}
\end{equation}

In our system, the repercussion of making a missing error is much more severe than making a false alarm error. For example, we aim to make fewer mistakes in detecting no gun when a concealed gun is present, rather than detecting some other object as a gun. Since detecting a regular person erroneously as a gun holder results in a small check to keep people safe while letting an actual gun holder go undetected might lead to catastrophic outcomes.

\subsection{Single-shot Prediction Results}
After training a single-shot prediction network in the COD system with the open carry dataset and concealed dataset, the performance of the system is tested with detection threshold $p_\R{thr}=0.5$ and the metrics stated above. With the single-frame cropped cube as input, this method is represented as \textbf{COD-single} and the corresponding evaluation results are shown in the first six columns of Table.~\ref{tab:res_single}. Overall, the COD-single method performs best to detect an open carry laptop with a false alarm of \SI{19.48}{\percent} and a missing of \SI{14.2}{\percent}. The system also performs well with detecting open carry phones and knives, with false alarm percentages of \SI{24.46}{\percent} and \SI{27.69}{\percent} respectively and missing percentages of \SI{25.93}{\percent} and \SI{26.04}{\percent} respectively. However, when it comes to concealed objects, all metrics present a lesser performance. The system on average misses detecting the object about \SI{8}{\percent} more often, which can prove to be harder for the system to get a concealed knife or gun detected. In terms of precision, false alarm, and F1, the system performance as a whole declines by about $12$-\SI{17}{\percent} when the object is concealed compared to the open carry case. That is probably because more variations regarding object position or perspectives and cover materials (e.g., clothes, bags) are introduced in the concealed cases.


\subsection{Multi-shot Decision Results}
With the trained single-shot model, the system performance of the multi-shot decision module is evaluated here by using 10 in-track cubes corresponding to the same pedestrian subject. The system with multi-shot decision module is named \textbf{COD-multi}. The metrics are then recalculated for COD-multi with same threshold 0.5 and the obtained results are shown in the last five columns of Table.~\ref{tab:res_single}. The results tell that adding a multi-shot decision module improves the performance of system for both concealed and open carry objects. On average, missing and false alarm rates decreased by approximately $5$-\SI{8}{\percent}, while on the other hand precision, recall, and F1 increased by roughly the same amount. Particularly, for openly carried objects, the precision increased by \SI{6}{\percent}, recall increased by \SI{6}{\percent}, and missing went down by \SI{6}{\percent}. Even when the object is concealed, the system's performance improved by about \SI{7}{\percent} across the board when comparing results to COD-single. Decreasing false alarms, missing, and improving precision are important to establish an effective system, and providing a multi-frame cube input does just that.

\subsection{Qualitative Results}
Ten test samples of the system actively working are shown in Fig.~\ref{fig:qual_res}. For each sample, we have the visualization of input cube as well as the RGB image of subjects holding or carrying objects and the detection results of the system. The probability provided by system to determine which objects it detects is printed onto the RGB image with different colors representing different objects. In the first two rows, all the objects are openly carried, while all the images in the third and fourth row are concealed objects. Fig.~\ref{fig:qual_res} shows a glimpse of the diversity of the data that was collected, in which the object was placed in different locations, such as a pocket, backpack, held to the side, or held in front, in addition to randomized walking patterns. Given those samples with diverse situations, our system adapts well and is able to detect the object successfully by outputting correct probability larger than threshold. On the whole, the detection probability of concealed object is a little bit smaller than that of open carry object, which agrees with above quantitive evaluation results.

\subsection{Ablation Study}
\highlighttext{For the backbone design, we proposed to concatenate the feature maps from different ResNet layers to form the multi-scale features in Section.~\ref{sec:sspn}. To study the effect of feature concatenation, we conducted an ablation experiment of solely using the feature of last layer (no feature concatenation) and gave the performance of concealed object detection with this setup in Table.~\ref{tab:ablation}. Comparing with the original results in Table.~\ref{tab:res_single}, the system without feature concatenation in backbone tends to own lower false alarm, higher missing rate, and lower F1, which demonstrates that the no-feature-concatenation setup is less sensitive to detecting objects than the feature-concatenation setup and thus the latter one is more competitive for fitting system requirements.}

\begin{table}[h]
\centering
\caption{Evaluation results for concealed object using COD-single method with no feature concatenation.}
\resizebox{0.5\textwidth}{!}{
\setlength\tabcolsep{3pt} 
\begin{tabular}{llcccc}
\toprule
\textbf{Object} & \textbf{Method} & \multicolumn{4}{c}{\textbf{Metric}}\\
\cmidrule(lr){3-6}
\textit{concealed:} & COD-single & precision & recall (missing) & false alarm & F1\\
(a) laptop & (no feature &  0.4987 & 0.6414 (0.3586) & 0.2860 & 0.5608  \\
(b) phone & concatena- &  0.3174 & 0.6992 (0.3008) & 0.323 & 0.4365 \\
(c) knife & tion) &  0.6008 & 0.6058 (0.3942) & 0.3949 & 0.6032 \\
(d) average & &  0.4723 & 0.6488 (0.3512) & 0.3347 & 0.5335\\
\bottomrule
\end{tabular}}
\label{tab:ablation}
\end{table}

\begin{figure*}[t]
\centering
\includegraphics[width=0.98\textwidth,clip]{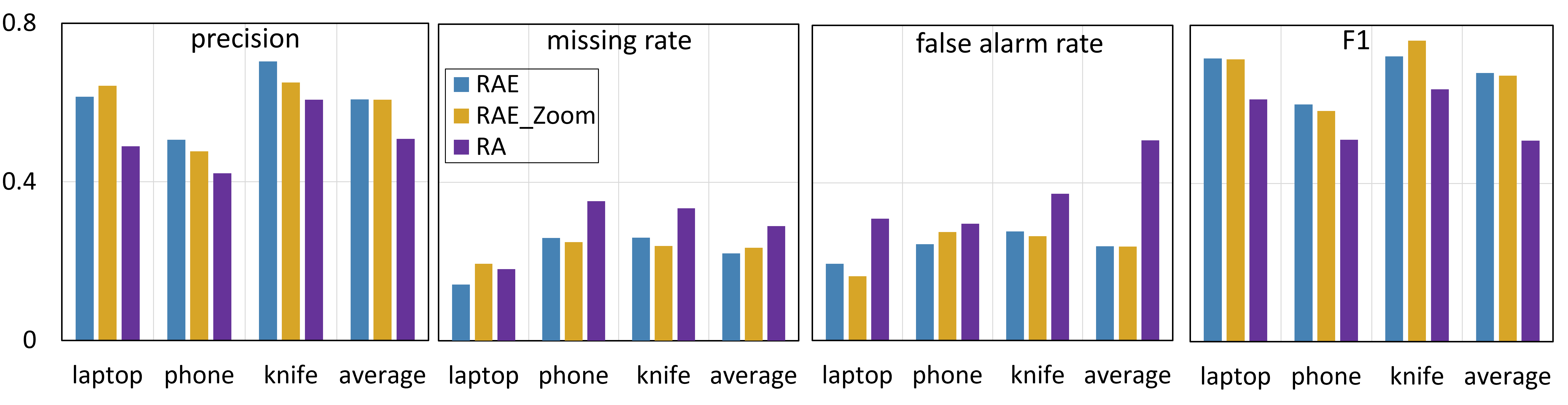}
\caption{Performance comparison for RAE, RAE-Zoom, and RA input. From left to right and up to down, four bar charts depict the precision, missing rate, false alarm rate, and F1 for \textit{open carry} object detection using COD-single model, respectively.}
\label{fig:res_input}
\end{figure*}

\begin{figure*}[t]
\centering
\includegraphics[width=0.98\textwidth,clip]{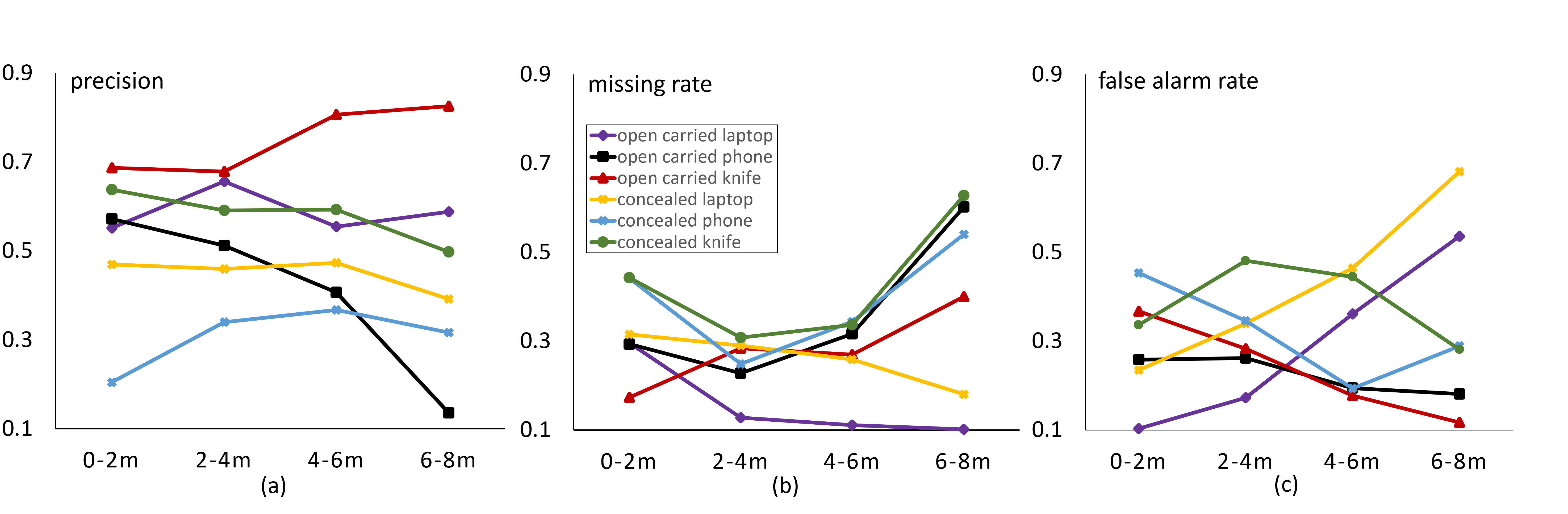}
\caption{Evaluation of detecting objects at different distances with COD-single model. From left to right, three subfigures present the precision, missing rate, and false alarm rate changes against distance, respectively.}
\label{fig:res_range}
\end{figure*}

\section{Discussion}
\label{sec:discuss}

\subsection{Influence of Input Data}
The selection of input data has a big impact on the system performance. Here, besides the range-azimuth-elevation (RAE) cube input, two more input formats are considered: \textit{range-azimuth (RA) cube} and \textit{RAE-Zoom cube}. 
\begin{itemize}
\item \textbf{RA cube}: RA is simply removing the elevation dimension from the data, keeping the range and azimuth angle dimension. By comparing the performance between RA cube and RAE cube input, a conclusion can be made about whether there is a benefit from the additional elevation dimension. To generate the RA cube, the Elevation FFT operation is removed from the preprocessing workflow shown in Fig.~\ref{fig:prepro}. Instead, we concatenate the Angle FFT processing results of different vertical RXs directly along a new dimension (i.e., similar to RGB channels). Since the input dimension is reduced, the 3D convolution layers in single-shot prediction network are accordingly replaced with 2D ones. The new model is trained from scratch using the method in Section~\ref{sec:imp-train}.

\item \textbf{RAE-Zoom cube}: RAE-Zoom takes an extra step of adjusting the RAE cube data by calibrating the input to a similar (cross-range) coverage plane. It is to manually handle the issues from polar-coordinate radar imaging where objects look small in the distance and big on the contrary. We solve this by a zooming in (out) operation that projects radar data to a fixed cross-range plane and then makes interpolation. The generated data is called RAE-Zoom cube, which will be input to the origin network for training and evaluation. 
\end{itemize}

Using the metrics defined in Section~\ref{sec:res_metric}, the performance of the COD-single method was re-evaluated using RA cube input and RAE-Zoom cube input. Fig.~\ref{fig:res_input} shows the evaluation results of those computations for open carry object detection using all three different input types mentioned earlier. When comparing the results of RAE-Zoom to that of RAE input, the numbers are very similar to the point where the zooming operation has no major impact on the results. Not including the zooming would be more beneficial since it will remove an extra step of computation when determining a presence of danger resulting in a faster response from the system. Comparing the results of RAE and RAE-Zoom to RA, the performance of the system is worse across all four metrics. On average, RA provides the lowest precision and F1, and the highest false alarm and missing rates. Based on these results, the additional elevation information in RAE and RAE-Zoom shows the capability of improving system performance across the board. Besides, the minimality of the improvement provided by the zooming operation in RAE-Zoom is outweighed by the speed of operating without it, which tells that additional zooming is unworthy and the proposed network can handle this variation inherently.

\subsection{Influence of Distance and Occlusion}

\textbf{System performance for different distances:} When detecting an object, distance plays an important role in determining the reflection amplitude of the object or subject thus affecting the effectiveness of the system. To evaluate how well the system works for different distances, the objects are divided into four groups - $0$-\SI{2}{m}, $2$-\SI{4}{m}, $4$-\SI{6}{m}, and $6$-\SI{8}{m} - according to their measured ranges. Fig.~\ref{fig:res_range} shows the precision, missing, and false alarm rates against the distance across six classes of objects: concealed and open carry phones, knives, and laptops. In Fig.~\ref{fig:res_range}(a), the plot for precision shows that on average the further away the object is, the lower the precision of the system. When an object is close to the radar it has a much larger reflection, which tends to make it easier for the system to detect, however being too close can lower the precision of the system. Meaning that there might be an optimal distance at which the system is most precise. Based on the plots in Fig.~\ref{fig:res_range}(a) the optimal distance for all the objects as a collective is between 2 to 6 meters.

Fig.~\ref{fig:res_range}(b) and \ref{fig:res_range}(c) show the plot for missing and false alarm rates with respect to the distance of the object. The plots reveal that the missing rate increases and false alarm decreases at distances greater than 6 meters except for open carry and concealed laptops. This is probably due to other objects having a smaller size in comparison to laptops, which drives the system to ignore them when further away. To get a better trade-off of false alarm and missing rate for long-distance objects, the detection threshold $p_\R{thr}$ can be adjusted accordingly.



\textbf{System performance for different occlusion:} An important observation to point towards is that the evaluation results of the system can differ based on the concealment condition of the object. Table.~\ref{tab:res_occlus} shows the evaluation results for two different experiments. In the first experiment, the training data consisted of all open carry objects, while in the second experiment the training data were all concealed objects. The testing data for both cases were concealed knives, which were further divided into three groups: knife in pocket, knife in bag 1, and knife in bag 2. Note that the precision and false alarm metric are omitted here since the test object being observed was only concealed knife thus precision is always 1 and false alarm is non-existent. Besides, F1 value here is greater than that in Table.~\ref{tab:res_single} because of the large precision in calculating F1.

From the comparison between two experiments, the system performed clearly much better when trained with concealed data with an increase in recall by about \SI{20}{\percent} or more for all three groups. This is due to the fact that in the presence of additional cover materials like a backpack, the radar receives extra reflections from them which may confuse the system only trained with open carry data. Therefore, training the system with concealed object data is important to maintain high detection performance. From Table.~\ref{tab:res_occlus}, even when training with all concealed data, the recall differs depending on the manner the object was concealed. For example, the system performed better when detecting knife in pocket than bag 1. This stresses the importance of diversifying the training data set to include multiple different concealment methods to improve generalization and performance.

\begin{table}[!ht]
\centering
\caption{Evaluation of detecting concealed knife with COD-single model under different occlusion condition.}
\resizebox{0.4\textwidth}{!}{
\setlength\tabcolsep{3pt} 
\begin{tabular}{llcccc}
\toprule
\textbf{Object} & \textbf{Method} & \multicolumn{2}{c}{\textbf{Metric}} \\
\cmidrule(lr){3-4}
 & COD-single & recall (missing) & F1  \\
\textit{concealed knife:} \\
(a) in pocket & (training with & 0.5038 (0.4962) & 0.6691 \\
(b) in bag 1 & all open carry & 0.3637 (0.6363) & 0.5331 \\
(c) in bag 2 & data) & 0.39 (0.61) & 0.5608 \\

\textit{concealed knife:} \\
(a) in pocket & (training with & 0.6947 (0.3053) & 0.8196 \\
(b) in bag 1 & all concealed & 0.6333 (0.3667) & 0.7754 \\
(c) in bag 2 & data) & 0.7282 (0.2718) & 0.8423 \\
\bottomrule
\end{tabular}}
\label{tab:res_occlus}
\end{table}

\begin{figure}[!t]
\centering
\includegraphics[width=0.5\textwidth,clip]{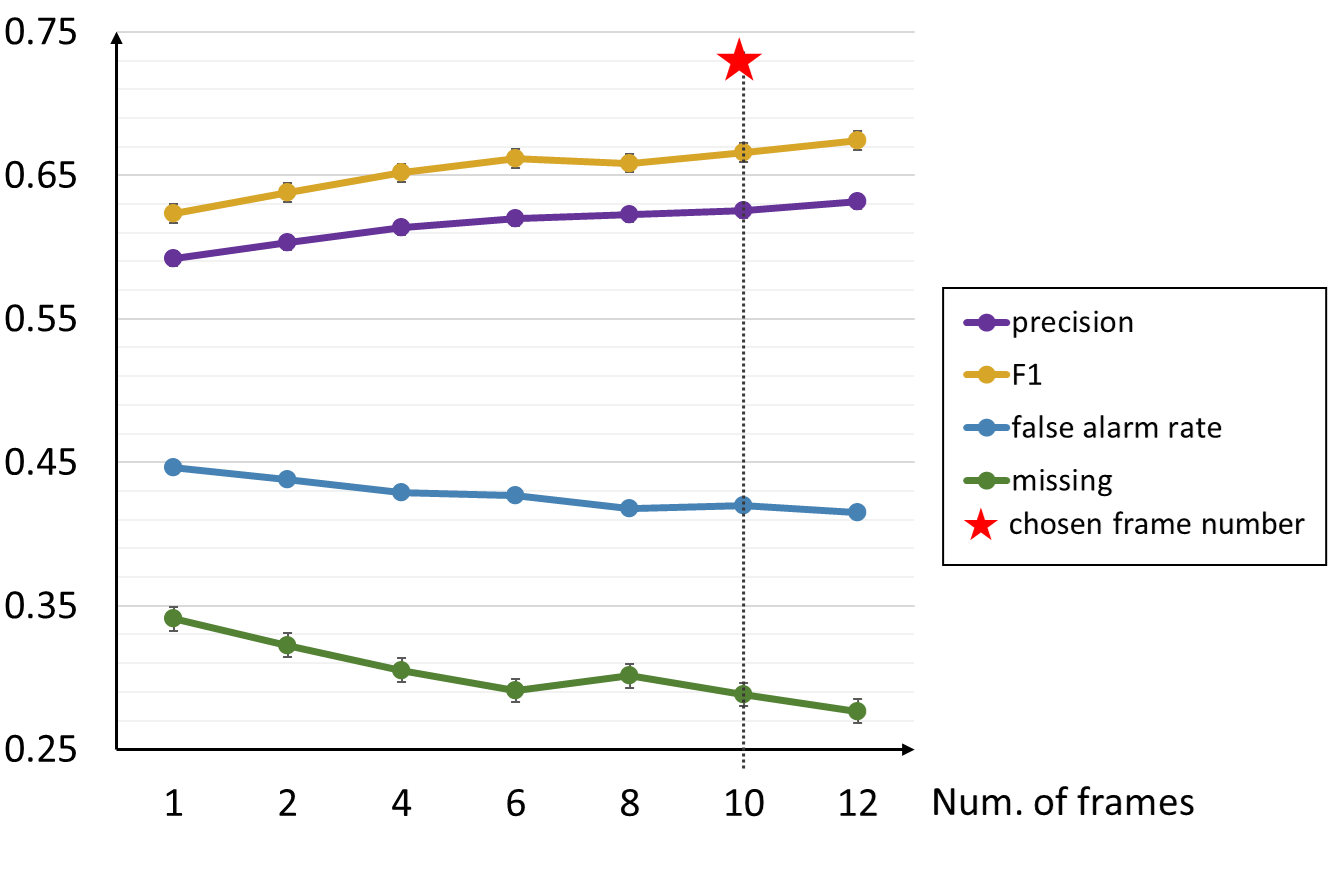}
\caption{Evaluation of detecting concealed knife with COD-multi model using varying-length input frames.}
\label{fig:res_frame}
\end{figure}

\subsection{Performance Trade-off by Adjusting Parameter Values}
\textbf{Number of input frames $N$:}
The number of frames used in the multi-shot decision module directly correlates with the efficacy of the system. To show it, we evaluate the performance of detecting concealed knives with varying-number of cropped cubes input and plot the results in Fig.~\ref{fig:res_frame}. It illustrates that using more frames increases the precision and F1, while also decreasing the false alarm and missing rates before they get saturated. However, using too many frames can be detrimental to the objective of our system, which is to provide security in real-time. Increasing the number of frames in the detection process will increase the required time to run the algorithm, hence delaying the result, which could prove to be fatal in a worst-case scenario \cite{Kowalski:19} \cite{s20061678}. Given the importance of real-time running, the plot and Table.~\ref{tab:run_time} help to select the ideal number of frames to improve the accuracy of the system, while maintaining an acceptable operating time. In this case, the selected ideal number of input frames is 10.

\begin{table}[!t]
\centering
\caption{Run time measurement for varying-length input}
\begin{tabular}{lcc}
\toprule  
& 1-frame input  & 10-frame input\\
\midrule  
run time per output & \SI{23.47}{ms} & \SI{215.63}{ms} \\
\bottomrule 
\label{tab:run_time}
\end{tabular}
\end{table}

\begin{figure}
\centering
\includegraphics[width=0.5\textwidth,clip]{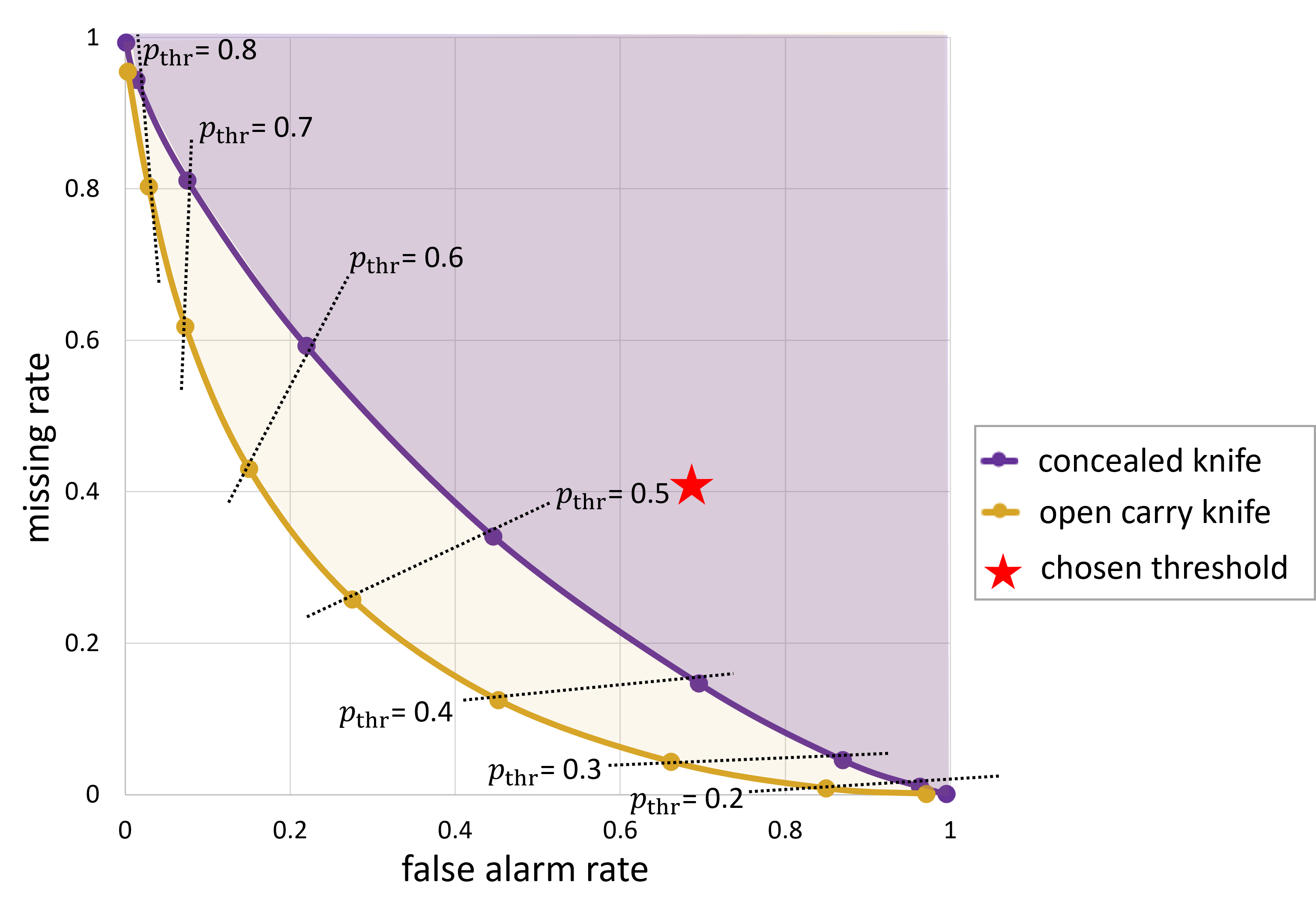}
\caption{Performance trade-off between false alarm and missing rate with varying detection probability threshold.}
\label{fig:res_prob}
\end{figure}

\textbf{Detection probability threshold $p_\R{thr}$:} This threshold is placed at the end of the system in order to identify a detection with a prediction probability greater than the threshold. Varying the detection threshold causes a performance trade-off between false alarm rate and missing rate as shown in Fig.~\ref{fig:res_prob}, where two missing-against-false alarm curves are plotted for concealed and open carry knives respectively by testing COD-single model with different threshold values. From Fig.~\ref{fig:res_prob}, it is easy to find a threshold to minimize missing rate with the price of increasing false alarm rate (vice versa). Aiming to achieve the best balance, we select the inflection point in both curves which provides the lowest combination of false alarm rate and missing rate. For both curves, that occurs at the threshold of 0.5.

\subsection{Comparison with State-of-the-art Methods}
\highlighttext{In Table~\ref{tab:compare}, a rough comparison was made between the proposed system and the state-of-the-art methods \cite{8650148, 8628238, 9269991, 9211448} in terms of objective, data type, running time and performance. Since the methods had quite different objective and input data type, we were unable to make a fair comparison by re-implementing them on our dataset and we only used the evaluation results and running time mentioned in papers. From the results, \textit{Bazgir etc.} \cite{9211448} shows good classification performance of walking with/without riffle, which however required the micro-Doppler images lasting more than \SI{6}{s} to capture the unique gait signatures and was uncertain about the viability of multiple object classification.  \textit{Wang etc.} \cite{8650148}, \textit{Liu etc.} \cite{8628238}, \textit{Wang etc.} \cite{9269991} implemented image processing and computer vision techniques on the high-resolution radiographic-like images generated by mmWave scanner, which brought less than \SI{10}{\percent} false alarm and missing rate, but required a long scanning time for generating those images. On the contrary, our COD system took extremely fuzzy RAE images as input and showed relatively acceptable performance for detecting three concealed objects with super fast processing speed \SI{215.63}{ms}. Despite rough comparison, we show that current COD system is 
with the potential of low-cost real-time dangerous substances screening as well as the urge of continued improvement.} 

\begin{table}[!h]
\centering
\caption{Comparison with state-of-the-art methods in terms of objective, data type, running time and performance.}
\resizebox{0.5\textwidth}{!}{
\setlength\tabcolsep{3pt} 
\begin{tabular}{lllcl}
\toprule  
\textbf{Method} & \textbf{Objective} & \textbf{Data type} & \textbf{Running time} & \textbf{Performance} \\
\midrule  
 \textbf{COD (ours)} & phone, laptop, & \SI{77}{GHz} &\SI{215.63}{ms} &  false alarm \SI{33.84}{\percent}\\
 & knife, others & radar & & missing \SI{23.85}{\percent}\\
 \midrule
 \textit{Bazgir etc.} & w/wo riffle & RF radar & $>$ \SI{6}{s} & precision \SI{94.5}{\percent}\\
  \cite{9211448} & & & & missing \SI{9.35}{\percent}\\
 \midrule
 \textit{Liu etc.} & w/wo object & mmWave & -- & false alarm \SI{9.73}{\percent} \\
 \cite{8628238} & & scanner & & missing \SI{9.95}{\percent}\\
 \midrule
 \textit{Wang etc.} & gun, knife & mmWave & -- & precision \SI{69.78}{\percent} \\
 \cite{9269991} & & scanner\\
  \midrule
 \textit{Wang etc.} & w/wo object & mmWave & -- & false alarm \SI{9.87}{\percent} \\
 \cite{8650148} & & scanner & & missing \SI{8.34}{\percent}\\
\bottomrule 
\label{tab:compare}
\end{tabular}}
\end{table}

\subsection{Strength and Limitation}
The proposed \SI{77}{GHz} mmWave radar-based COD system can detect three classes of objects - laptop, phone, and knife - under both open carry and concealed cases. Compared to the camera-based security system, it works in the low-light and object blocking or occlusion scenarios and comes without any privacy concerns. When compared to current security inspection techniques, e.g., X-ray, mmWave imaging, it lowers the requirements for taking an image with fixed posture or position and greatly reduces the processing time. Our system is capable of generating real-time object detection output for both moving and static pedestrian subject, by proper signal preprocessing and deep learning model. The time efficiency improvement enables more flexibility and higher passenger throughput rates, however, may also cause accuracy loss as a trade-off. From the observation that the multi-shot decision module brings complete performance improvement, it is promising to further explore methods of utilizing multiple frames in the future.

For limitation, the proposed COD system requires the sensor and target to be in relatively close proximity (e.g., $<\SI{6}{m}$) to keep good detection capability, just like all other approaches. Besides, the experiments are limited to the single carried object situation in this paper, which should be further extended to more complicated scenarios, e.g., a subject holding multiple objects, multiple subjects holding different objects. Due to the difficulty of capturing data in such circumstances shortly, this must be left for future work.


\section{Conclusion \& Future Work}
\highlighttext{In this paper, we focused on the relatively unexplored area of carried objects detection with low-cost \SI{77}{GHz} mmWave radar to foster the next-generation human safety inspection techniques. The proposed COD system is capable of real-time detection of three classes of objects - laptop, phone, and knife - under open carry cases and concealed cases where objects are hidden with clothes or bags. This system would be the very first baseline for other future works targeting carried object detection using \SI{77}{GHz} radar and the analysis of effects of different parameters provides important physics-based insights into problem-solving. For future work, more experiments and evaluations of testing system efficiency would be performed and more exploration of utilizing multiple frames would be continued to push the performance limit.}

\bibliographystyle{IEEEtran}
\bibliography{bibtex}

\appendix
The details of the backbone layers of the proposed system are shown in Table.~\ref{tab:layers}.
\begin{table}[!h]
\centering
\caption{Details of backbone layers.}
\resizebox{0.47 \textwidth}{!}{
\renewcommand{\arraystretch}{1.2}
\begin{tabular}{c|c|c}
\hline Layer name & \multicolumn{2}{c}{conv1$\underline{\enskip}$x} \\
Filters & \multicolumn{2}{c}{$3\times3\times3$} \\

Output size & \multicolumn{2}{c}{$24\times24\times10$} \\

\hline Layer name & conv2$\underline{\enskip}$x & conv3$\underline{\enskip}$x \\

Filters & $\left[\begin{array}{c}1 \times 1 \times 1,64 \\ 3 \times 3 \times 3,64 \\ 1 \times 1 \times 1,256\end{array}\right] \times 3$ & $\left[\begin{array}{c}1 \times 1 \times 1,128 \\ 3 \times 3 \times 3,128 \\ 1 \times 1 \times 1,512\end{array}\right] \times 4$ \\

Output size & $12 \times 12 \times 5$ & $6 \times 6 \times 3$ \\

\hline Layer name & conv4$\underline{\enskip}$x & conv5$\underline{\enskip}$x \\

Filters & $\left[\begin{array}{c}1 \times 1 \times 1,256 \\ 3 \times 3 \times 3,256 \\ 1 \times 1 \times 1,1024\end{array}\right] \times 6$ & $\left[\begin{array}{c}1 \times 1 \times 1,512 \\ 3 \times 3 \times 3,512 \\ 1 \times 1 \times 1,2048\end{array}\right] \times 3$ \\

Output size & $3 \times 3 \times 2$  & $2 \times 2 \times 1$ \\
\hline
\end{tabular}}
\label{tab:layers}
\end{table}

\end{document}